\documentclass{solarphysics}

\usepackage[hyperref,optionalrh,solaromanenum]{spr-sola-addons} 
\usepackage{enumitem}
\usepackage{aas_macros}
\usepackage{graphicx}  
\usepackage[figuresright]{rotating}
\usepackage{caption}

\usepackage{xcolor}                       

\usepackage[T2A,T1]{fontenc}

\newcommand{\ion}[2]{#1\,\uppercase\expandafter{\romannumeral #2\relax}}

\begin{document}

\begin{article}

\begin{opening}

\title{My Rewarding Life in Science}

%
\author[addressref={},corref,email={mjberdy@mac.com}]{\inits{A. P.}\fnm{Andrew}~\lnm{Skumanich}}

%
\runningauthor{A. Skumanich}
\runningtitle{Skumanich Memoire}


\begin{abstract}
This memoire covers my life history starting with my family’s
background and their immigration to the US. It continues with my
childhood, my early education, and my introduction to science. It then
covers my professional research career included a variety of institutions
and areas of Physics ending ultimately in Solar Physics.
\end{abstract}

%

\end{opening}

%
\section{Skumanich Family Background}

My father Petro Skumanich was born to a farming family in P\v{c}olina (the
town of Bees) about 1894 in the Zemplen district of the Austro-Hungarian
empire (now Slovakia). My mother Mariya Scrip was born about 1895 in
\v{C}ukalovtse (Chukalovtse) the next village higher up Northward from
P\v{c}olina. These towns and villages were on the South side of Carpathian
Mountains whose spine or crest divided Polish-Rusyn Galicia from
Austro-Hungary. The Skumanich and Scrip families owned their farms, as
did many others.

Hungarian was the primary language of the Zemplen administration. My
father could speak and read Hungarian. His mother's tongue at home was
Rusyn, an Eastern Slav dialect. He learned to read and write at the
local (and only) Uniate (Greek Catholic or Byzantine) Church school. He
also knew how to read (and presumably speak) Polish, Ukrainian, and
Slovak. He could read and understand Church Slavonic. Much of this must
have come from his studies to be a priest. This ambition was never
realized due to an unfortunate accident in which he was responsible for
the death of a village youth. This crime caused the priesthood to be
closed to him.

My mother had a 3rd-grade education. She could  slowly read and write Rusyn. She often had me write her postal letters in Rusyn and later in
English when I learned how to read and write English. Nonetheless, she
was quite competent in the New World.

With the outbreak of WWI, my father was conscripted into the
Austro-Hungarian army to fight on the Eastern front. Early in the war he
was exposed to poison gas due to a `friendly fire' incident. He was
captured by the Russians soon after and spent the remainder of the war
in a Ukrainian commune, a farming cooperative, escaping only after
rising through the administrative ranks.

After the war my father decided to join his older brother who had
immigrated, before the war, to Cleveland, OH. My father wanted to
buildup a `nest egg' and return to P\v{c}olina to buy up land to add to his
inheritance.

His brother was working for a railroad company and managed to get my
father a good job in the company as a switch-yard locomotive driver.
Unfortunately, somewhat later, he had an alcohol-related accident which disabled the
locomotive. To avoid prosecution, he fled to a mining town, Blackman's
Patch in Wilkes-Barre Township, neighboring the city of Wilkes-Barre,
Pennsylvania. There were countrymen from my father's village living in
the Patch drawn to work in the Anthracite mines.

My father passed the requirements for obtaining Mining Papers
(engineering certificate). That allowed him to direct the work of
laborers and to decide which coal veins to attempt to open, where to set
dynamite charges to break up the veins and to direct their removal. He
was also able to make tools that he required to facilitate his work.

Ultimately he decided to stay in the US and had my mother join him in
late December 1928. It must have been a joyful reunion. I was born at
home in the Patch just 9 months later on Oct 5, 1929. I was my mother's
second pregnancy, this time in the US, and the first to survive.

My father's life as a miner was a difficult one. He would come home
black with coal dust and wash in a portable galvanized circular floor
tub. My mother would heat water on the kitchen stove and pour it over
him and then wash his back. It was not an easy bath. Given all the other
attendant mining ill effects, I vowed that mining was not for me.

\section{Childhood and
Education}

The language of discourse in my parents' home was always Rusyn, a
western dialect of Russian, also identified as Ruthenian or
Carpatho-Russian. My literacy training in Rusyn was at home. I started
Primary school with a smattering of Street English that was quickly
replaced. My Russian literacy was developed in a contemporaneous Russian
Church school along with Russian and Church Slavonic vocabularies. The
latter was used for liturgical purposes. On rainy Sundays my parents
celebrated the Liturgy at home with my father as the priest and cantor
with my mother, brother, and me as the choir. My earliest ambition was
to become a priest. This ambition was ultimately quashed when my father
resigned from the Church after a bitter row with a new ambitious priest.

In my childhood years I was drawn to the comics. My mother had a boarder
that subscribed to a Sunday paper. My younger brother and I would
surreptitiously remove the comics, carefully read them and return them
to the `undisturbed' paper.
The comic that had a lasting effect on my love of Art and History was
Hal Foster's Prince Valliant. An adventure strip, set within the Middle
Ages around knighthood, with a line art that pleased the eye with its
verisimilitude and richness of color. My new ambition was to learn to
draw as well as Foster. I bought books that offered self-taught courses
in drawing.

I had the pleasure of an art accomplishment in a primary school
incident. I drew a free-hand copy of the fa\c{c}ade of my school building
and showed it to my class teacher. She liked it and had it printed in
the school bulletin. My art was on display!

\section{Introduction to
Science}

The singular event which launched me into the study of Astronomy and
ultimately Physics was a brilliant red (oxygen) nighttime aurora in
August of 1939.
My mother called me outdoors to view a bright sky that scintillated with
blood-red colors. I was awestruck. She claimed it was an omen
foretelling a bloody war. Indeed Germany invaded Poland a month later,
September 1939, of course a coincidence.

The next night I went out to see if the event would be visible again.
No, the sky was dark, but I noticed that there were points of light that
I had never noticed before. I asked about them of my grade school
teacher she directed me to the school librarian who labeled them as
`stars.' She began my study of the stars by bringing me various books
from the public library.

MypPrimary school years were involved in learning the constellations and
their associated classical myths, the names of the visible stars, and
other night-sky phenomena. I attempted to make a telescope with Edmunds
Scientific lenses and tubes but never finished the project due to a
lack of tools.

At one Christmas I was given a chemistry set from which I learned of the
chemical elements and chemical reactions. One of which was making a
mixture of powdered zinc and sulphur which when heated would `explode'
with a flash and much fumes, a fabulous experiment.

Early in my secondary school experience I was enthralled by Edwin
Hubble's ``The Realm of the Nebulae.'' In the equation relating absolute
magnitudes to luminosities, I found the mystifying word `log.' When I
asked the school math teacher to explain what was meant he brushed me
off with the answer that I would learn about `logs' in next year's
course. I was disappointed but realized that I had to learn mathematics
in order to understand the language of science.

Later I joined The American Meteor Society as a student observer. This
entailed reporting meteor number counts and entering paths onto
celestial maps of the night sky. I even convinced a group of my high
school peers, while we were doing contract farm labor in the Finger
Lakes (NY) area, to help me with observations of the 1947 August Perseid
Shower. We brought out our mattresses and arranged them in a circle with
our heads at the center. Someone kept count, the others would call out
an event and I would plot the path. Alas my `crew', exhausted by the
day's labor, fell asleep before the peak of the shower. I labored on
alone.

\section{Higher Education}

I had held a variety of jobs during my final years in secondary school,
initially for family finances and finally for college tuition. By the
summer of 1947 my father had succumbed to anthrosilicosis (black lung)
from coal mine dust. Thus my mother was thrust upon public welfare.

The summary nature of my secondary school physics courses inspired me to
pursue additional more detailed studies. For economic reasons I started
at the local Wilkes College (initially Bucknell Jr. College) in
Wilkes-Barre and entered into a two-year Physics Certificate program in
the Fall of 1947. I supported my school expenses by working in the
college bookstore as well as by commercial jobs.

I met three other physics matriculants from the suburbs, William (Bill)
Holak, Angelo Campanella, and Fred Bellas. We became life-long friends.
Holak's ethnic and religious background was essentially identical to
mine, so our bond became familial.

Near the end of the first semester of our sophomore year (last year at
Wilkes) Holak, Campanella and I learned that Penn State would not insure
our transfer there for the fall 1949 semester. However we could and did
transfer in the second half of our sophomore year at Wilkes.

After settling down at Penn State I learned that I had to pay a
matriculation and dorm fee, money that I did not have. I was on the
verge of being dropped when Campanella's father, a grocery store
merchant, paid my fees. I have been forever grateful for his kind help.

During my junior year (1949-50) at Penn State I was awarded a one-year
State Senatorial scholarship. Subsequently in 1950 I started to work at
menial jobs. Later I was awarded a summer research assistantship in the
department to work with one of the professors to reduce molecular gas phase data to allow the determination of Van der Waals parameters. In my senior year I was hired by the technical firm HRB,\footnote{Named after the department Physics professors, Haller, Raymond, and Brown.}
where one of my duties was the numerical inversion of fairly large matrices associated with infrared images with the then newly available desktop Wang electro-mechanical calculator.

Although there were no Astronomy courses at Penn State, there was an
Astronomy Club, Alpha Nu, in the Physics department sponsored and led by
Prof.\ Yeagley.
One of our activities was to make ourselves our own 4-inch reflecting
telescope. To this end we ground our own parabolic mirror and then
bought our own supporting tube. Unfortunately for me I left in the
middle of the project and seldom had the time or an available mechanical
shop to finish my project.

A second life changing event was a chance encounter in my final semester
(1951) with one of my Physics professors, Richard Stoner (Princeton PhD
in Physics) who, when I told him I was planning to do graduate study in
Solid State physics at Case (Western Reserve) University, insisted that
I apply to Princeton instead. He helped me with the forms. The Physics
department student roster was full so my late application was forwarded
to the Astronomy department where I was accepted (one of only 2 admitted
annually) and awarded a Higgins Fellowship. Now I had the opportunity of
studying both fields, Physics and Astrophysics.

I entered Princeton in the fall of 1951 with a BS in Physics and left in
the fall of 1954 with a PhD in ``astronomia vestiganda.'' The scientific
and social milieu at the Department and its associated Observatory
during this period is well described by \cite{2000ARA&A..38....1O}.

The department had a two-year cycle of the standard courses that allowed
much time for Physics courses. Thus, I rounded out my Physics knowledge
with a variety of courses such as Quantum Mechanics by R.~H.\ Dicke (well
known for his work in Cosmology), Thermodynamics by H.\ Snyder
(Gravitational collapse singularity with Oppenheimer) and Mathematical
Physics by P.~G.\ Bergmann.

The departmental rule was to assign graduate assistants to work with
each academic member. In my first semester (1951) I was assigned to
Lyman Spitzer to help determine the internal motions within interstellar
clouds. It appeared in the Astrophysical Journal, and was my first
scientific paper \citep{1952ApJ...116..452S}.

\begin{sidewaysfigure}
    \centering
    \captionsetup{font=footnotesize}
\includegraphics[scale=0.55]{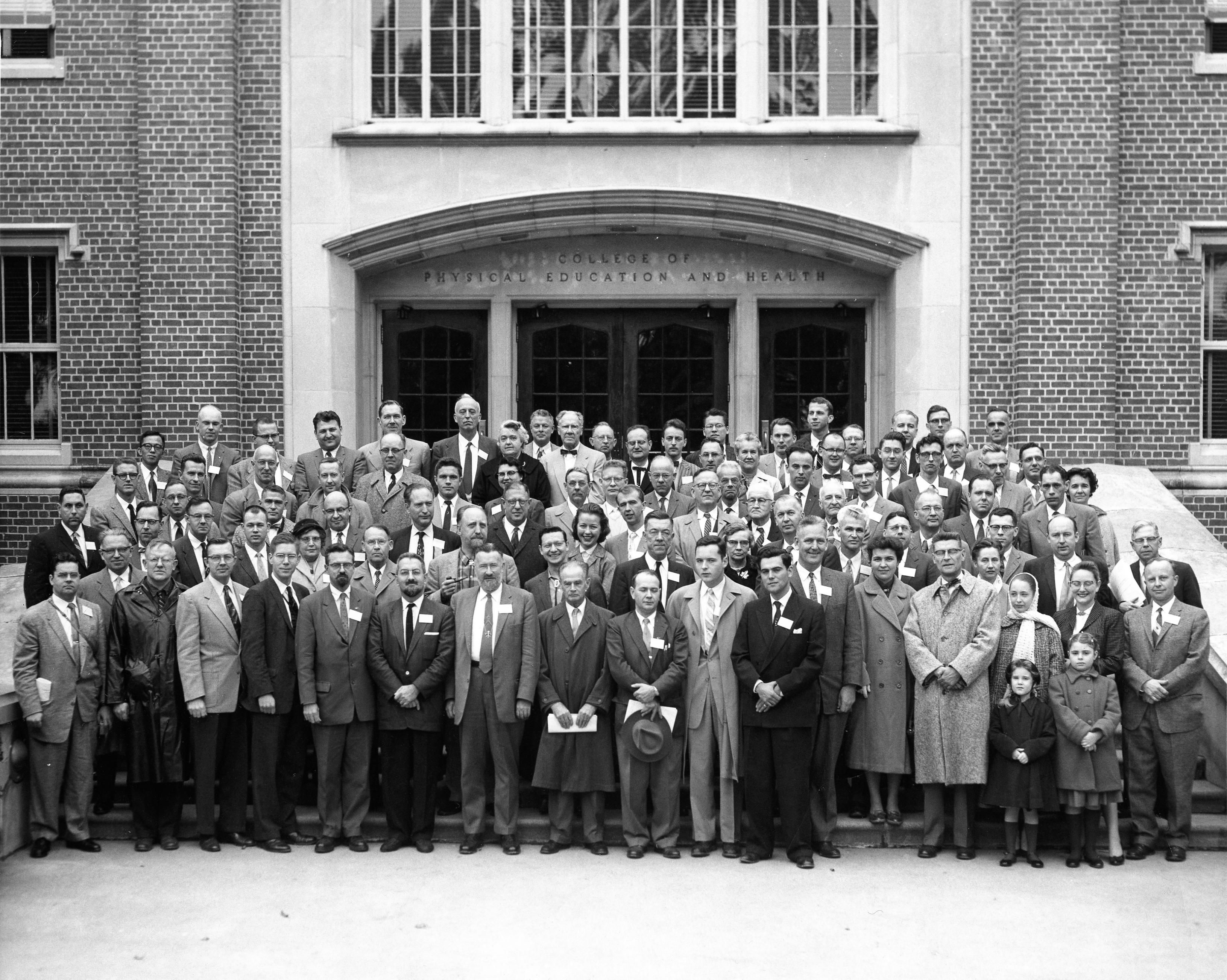}
    \caption{Attendees of University of Florida 1958 AAS Meeting. Among them: F.~K.\ Edmondson (AAS Secretary, front row, 6th from left); A.~N.\ Vyssotsky (Leander McCormick, front row, 7th from left); A.\ Skumanich (LASL,, 2nd row, head tilted slightly right, just behind Vyssotsky); A.~N.\ Cox (LASL, 4th row, just to the left of the woman above Skumanich and Vyssotsky); George Field (Harvard, right back row, tallest figure); J. Bev Oke (CalTech, in the next column, left of Field but two heads down).}
    \label{fig:1958 AAS}
\end{sidewaysfigure}

My next assignment, spring 1952, was with Uco van Wijk to help rectify
the light curve of the eclipsing binary GL Carinae so as to determine
the orbital properties of the star. The method of \ ``rectification of
light-curves'' was published in 1946 by H.~N.\ Russell, formerly head of
the Department. The analysis was continued by J.\ Rogerson, my
co-matriculant. The results appeared in \cite{1955AJ.....60...95V}.

In the summer 1952 I was sent to work with Prof. A.~N.\ Vyssotsky at the University\ of Virginia, Leander McCormick Observatory in Charlottesville, VA. The
purpose was to determine if there was a difference in the dispersion of the ‘proper’ motions of two spectroscopically
distinct Main Sequence dwarf G and K field stars.
The two groups were identified by a difference in their spectral G-band
or molecular CH composition. This band is an indirect measure of the
carbon ash, derived from the nuclear `burn' region, which increases with
age.
A kinematic age effect was indeed detected, with the weak G-band stars
having a smaller dispersion than the strong G-band stars \citep{1953AJ.....58...96V}.

Note that \cite{1951ApJ...114..385S} as well as \cite{1952ApJ...116..164O} had showed that stellar collisions with
interstellar clouds would `heat up' stars, i.e., increase the dispersion
of their `proper' velocities.

The Leander McCormick Observatory was a large structure similar to
Princeton's with offices, classrooms and a visitor living quarters. An
attached building housed a 26-inch refractor for use in parallax and
proper motion studies. It was sited on a hill overlooking the city. I
lived in the visitor quarters.

I found Vyssotsky to be a very knowledgeable and warm mentor whose
Russian cultural background enriched my Slavic sentiments. He had a rich
collection of talents. One of his talents, which he helped me to
develop, was to find a particular page in a book without a search, i.e.
at first pass.

I learned from Vyssotsky that he had been a staff astronomer at Pulkovo
Observatory (St. Petersburg) when the Russian revolution broke out. He
became a Captain in the White Army and later, when the Reds won, escaped
via Crimea to, ultimately, the University of Virginia. He later learned
that the subsequent observatory administration erased his name from all
the Observatory records. Years later, when I was at a conference on
magnetic fields and Stokes polarization at Pulkovo, I mentioned this
history to the local staff and, when I asked about his name being
reinstated, I was greeted with a culpable silence.
I fondly remember one of my last occasions to meet with him, at the 1958 AAS meeting at University of Florida (Figure~\ref{fig:1958 AAS}).

My final assignment in the fall of 1952 was to Donald (Don) E.\
Osterbrock, a visiting Post-Doc. I was to calculate, using a tabletop
machine, a parametric sequence of interior models of red dwarf stars \citep{1953ApJ...118..529O}. Thus I
learned to construct stellar interiors from Osterbrock. His course in
Stellar Atmospheres introduced me to the field of radiative transfer.

It is interesting to note that at this time (Spring 1953) the Princeton
electronic computer (ENIAC at the Institute for Advanced Study) was
online and was being used by Schwarzschild \citep{2000ARA&A..38....1O} to solve
the static hydrodynamic equations (stellar structure) with hydrogen
fusion, while at the same time Foster Evans (Los Alamos Scientific
Laboratory) was at Princeton solving the time-dependent hydrodynamic
equations with nuclear reactions (basically Supernova-like equations) in
his study of the feasibility of the ``Super'' \cite[thermonuclear H-bomb;][]{Evans1996}.

My third life-changing event was in the spring of 1953 when I, by pure
chance, met Bill's first cousin Mary J.\ Berdy. It was while Bill and I
were standing on a street corner waiting for a bus to downtown
Wilkes-Barre. She stopped for the red light, saw him and offered him a
ride. We spent the evening with her. I was completely charmed. She
became my wife in the fall of 1955.

The University required proficiency in two languages for the Doctoral
program. My German exam was given by Schwarzschild who complimented me
on my native-like use of the language. My Russian exam was given by John
Turkevich, Eugene Higgins Professor in Chemistry, which I passed easily.

Turkevich proposed that I and another graduate student, Alexander
Tulinsky, translate the Russian 
{\fontencoding{T2A}\selectfont   
Журнал Зкспериментальной и Теоретической Физики 
\fontencoding{T1}\selectfont}
(Journal of Experimental and Theoretical Physics) cover to cover, 
which he would submit to NSF to initiate a translation program of said
journal. He was awarded the contract to this effect with us as
translators. This started the Russian translation program at NSF. This
translation program was transferred to AIP and ultimately to Consultants
Bureau as publisher. I remained a translator through these changes `till
the late 1950's.

\begin{figure}
    \centering
    \includegraphics[width=.99\textwidth]{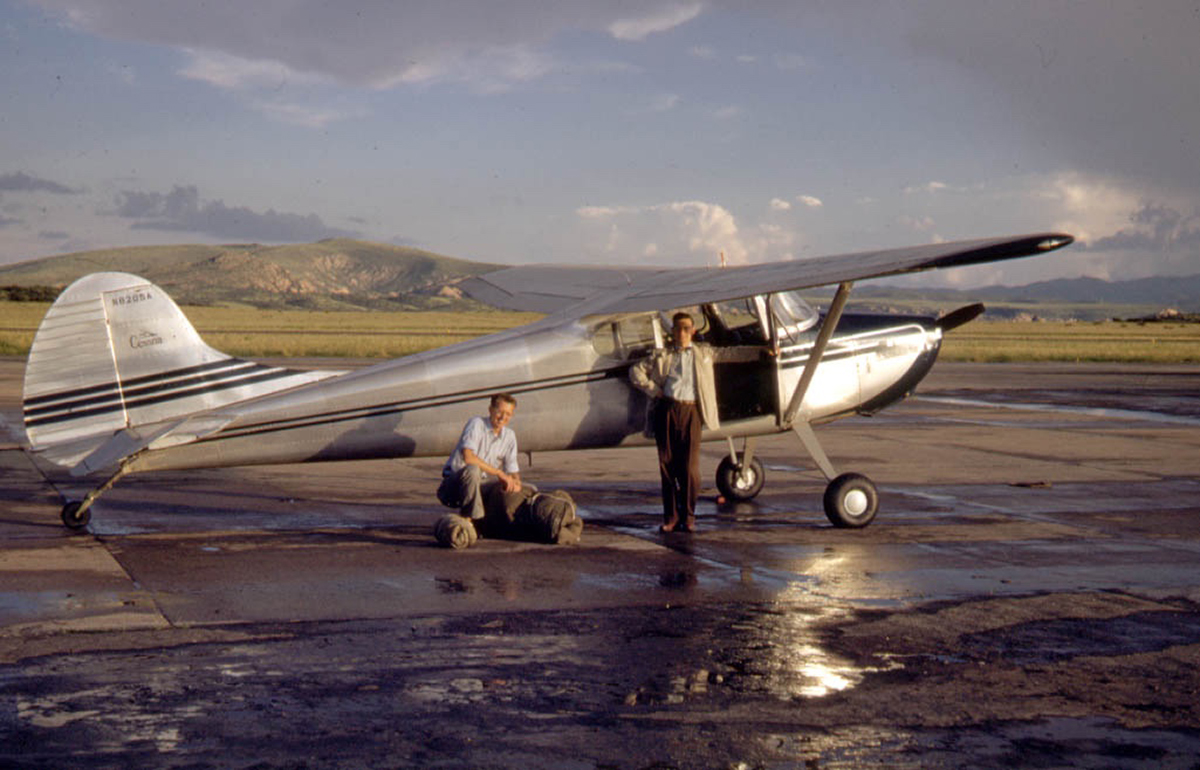}
    \caption{Bill Livingston  (left), Skumanich (right) and Dale Vrabec (photographer) have arrived at the Prescott, AZ airport}
    \label{fig:Cessna}
\end{figure}

\begin{figure}
    \centering
    \includegraphics[width=.99\textwidth]{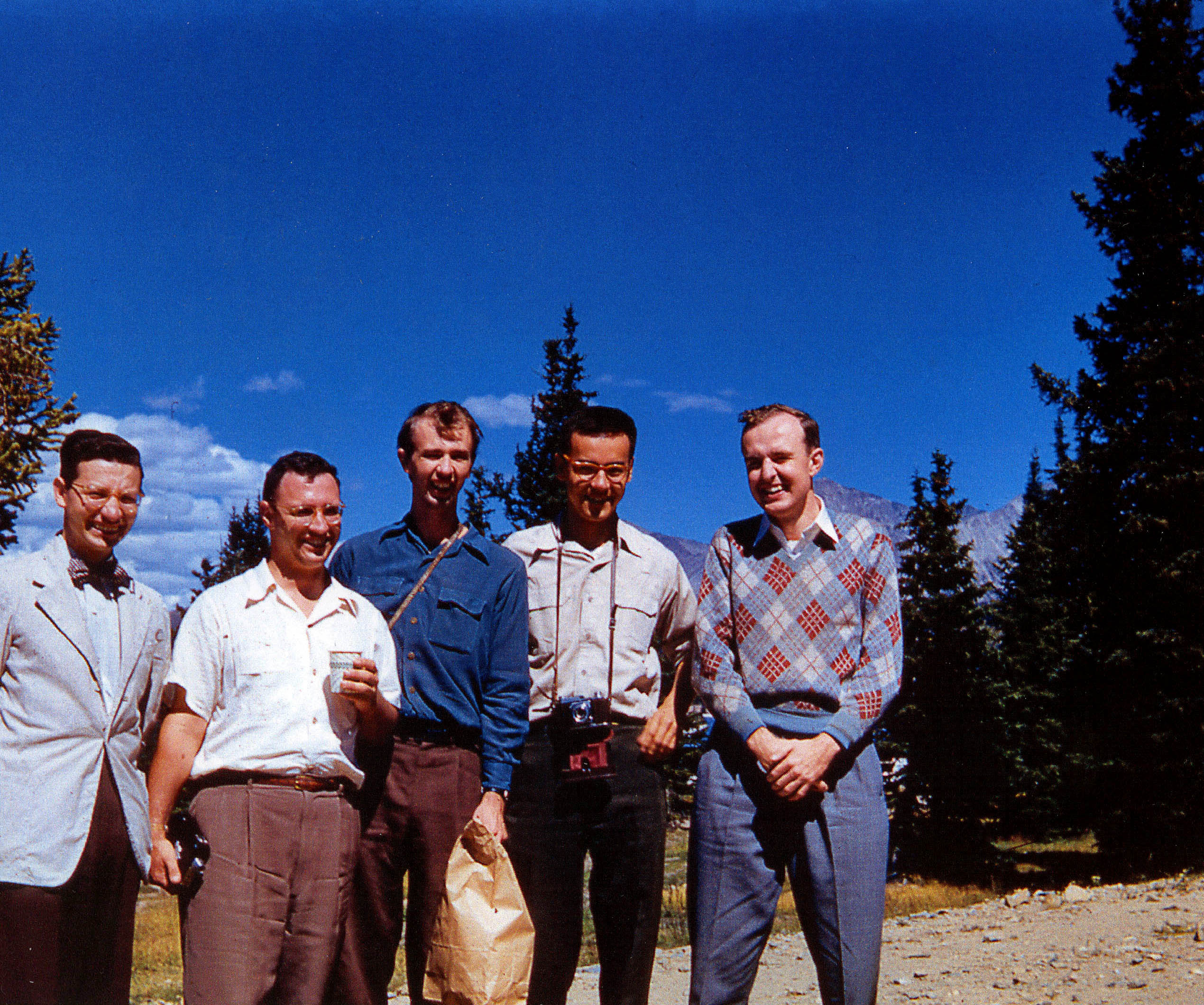}
    \caption{Skumanich (Princeton Univ., far left) with four young colleagues on the Top of the World (Loveland
Pass) to attend the 1953 AAS-Meeting in Boulder, CO. From the left: (unidentified), Art Cox (Indiana Univ.),
Dale Vrabec (CalTech), and Allan Sandage (Mt. Wilson).}
    \label{fig:1953 AAS}
\end{figure}

I started on my thesis in the Summer of 1953 with a three-month visit to
the Mt Wilson Observatory in Pasadena, CA, under the tutelage and guest
investigator privileges of Martin Schwarzschild who became my thesis advisor.
The trip protocol was part of the Department process for thesis
candidate to obtain their observational material or observational
experience. I traveled out there on the Atchison Topeka and Santa Fe
Railroad, a romantic introduction to the American West.

In my case Martin had suggested I look into the bright-dark symmetry or
lack thereof in granulation in order to investigate the nature of the
convection. The extant white-light images archived at the Observatory as
well as the appropriate analysis devices were available to me.

My analysis of the bright-dark wings of the lateral intensity
distribution indicated no significant areal asymmetry. I found a null
result as far as the nature of the convective turbulent heat flow \citep{1955ApJ...121..404S}.

During my Pasadena stay I became friends with William (Bill) Livingston
and Dale Vrabec, grad students at CalTech. Given the imminent Summer
1953 AAS meeting in Boulder, CO, Vrabec, an experienced pilot, decided
to fly out to Boulder with a loaner Cessna and invited Bill and me to
join him. It was a remarkable experience  (Figures~\ref{fig:Cessna} and \ref{fig:1953 AAS}).

On one of several Pasadena outings, I raised the question with Martin as
to what might fix the dominant granulation size. I asked whether this
was known. He knew of no such studies, so he suggested I look into the
problem as part of my thesis.

Thus I turned to the study of convective flow in a density-stratified
atmosphere that is thermally unstable. The previous study was Rayleigh's
(\citeyear{1916Natur..97..318R}) who considered a homogeneous atmosphere. My results, for an
adiabatic flow, indicated that the smaller the perturbation compared to
the thickness of the unstable zone the more unstable the mode \citep{1955ApJ...121..408S}. I found that for larger perturbations Rayleigh's
solutions were confirmed. My analysis benefited from discussions with
Arnulf Schl\"uter (Dr.\ Rer.\ Nat., Rheinische Friedrich-Wilhelms
University, Bonn, 1947), a Department Visiting Fellow, and with Richard H\"arm,
Schwarzschild's research assistant, in regard to the numerical
integration method.

My thesis stimulated a decade of further studies by others with one or
another of my assumptions removed (see, e.g., Table~1.[3])
These
studies also failed to produce a horizontal convective mode size that
might explain the observed solar granulation.

My thesis examination committee included the full Department staff,
Spitzer, Schwarzs{-}child, van Wijk and the venerable J.~Q.\ Stewart of
Russell, Dugan, and Stewart fame \citep{1926arya.book}. The procedure was going well until it
was Stewart's turn. He started by describing an experiment that
consisted of dropping a steel ball onto a plate sheet of glass and
ranking by size and counting the number of shards in any size group. I was
non-plussed when Stewart asked me whether I would discuss why the
frequency-rank distribution was given by a power law. I had no answer
and reacted angrily to what I thought was a non-thesis question. I was
embarrassed by my intemperate reaction and felt I had ruined my exam. It
seems my committee overlooked this small brouhaha and passed me after
all.

Looking back after so many years I still feel it was not a suitable
thesis question but nonetheless it posed an interesting problem. By now
this question must surely be resolved. It could have relevance in
determining the nature of the debris from the collision of two
asteroids. My diploma was formally issued Oct 22, 1954.

In the early winter of 1954, I began to think of employment prospects
for my post-graduate future. My first effort resulted in an interview at
the Army Ordnance Lab in Durham, NC. I took a train down to Durham to
interview for a position as a project scientist to review and select
research proposals from industry and universities. I would be in charge
of overseeing whatever research I approved. What I remember most of this
trip was the red clay of this southern state. I declined the position.

I also decided to visit the Winter 1954 meeting of the American Physical
Society (of which I was a member) to be held in New York City. At the
meeting I came across a professional booth staffed by Los Alamos
Scientific Laboratory (LASL) personnel presenting the science to be found at
the Lab. They were also advertising a position that required knowledge
of Spectroscopy and Radiative Transfer. I could certainly fill the
position so I applied then and there. This ultimately led to filling out
various forms to be submitted to the relevant individuals. After some
time I learned that I was a possible candidate and I was invited to come
to Los Alamos for a personal interview. I passed the interview so that
an extensive background investigative process, including a security
clearance investigation, began. I was offered a position in the weapons
test division.

In the meantime, I also sent my CV to Donald (Don) H.\ Menzel, whom I had met when he
visited the Department, for a Post-Doc position at Harvard. My thought
was to continue my convective flow analysis. Apparently Schwarzschild
had recommended me highly to Menzel who proposed to Walter Orr Roberts,
director of the High Altitude Observatory  in Boulder, CO, that I be
considered for a position as staff scientist. I was offered the
appointment to pursue solar physics at the Observatory.

By August 1954 I had two job offers to consider. I decided to accept the
Los Alamos position because it offered an interaction with very
knowledgeable academic physicists from a wide range of fields, a
Post-Doc experience. In addition it paid more that allowed me to
financially help my mother.

\section{Los Alamos Scientific Laboratory
(LASL)}

For my interview I had flown to Albuquerque and from there to Los Alamos
with Carr Airlines, a small carrier, that had a contract with the
University of California to serve that specific (classified) route. I
remember how appalled I was to see from the air a burnt out and
desiccated region. But after some time at LASL I fell in love with the
landscape. I finally understood why St.\ Anthony choose to live in a
desert.

Los Alamos was a small, closed city and mostly confined to two mesa
tops. The buildings, except for Fuller Lodge - a large log style two story
inn, were government style, wood sided, one or two story structures.
There was a water tower and a machine gun posted guard tower at the
fenced single entrance with a high wire fence around the entire site. It
was possible to access the town only with an official ID card, mine also
carried my Q-clearance (top secret) designation. I found the population
very friendly and well cultured.

When I arrived at Los Alamos in the Fall of 1954, I was assigned a
dormitory room in one of the temporary multi-room, two story barracks on
the North side of the main Canyon Rd. The town center was between Canyon on the North and Trinity Drive on the South. The lab was on the South side of Trinity overlooking a deep canyon. I walked to work to Building C, a one story wooden structure with an IBM punch card computer.

In 1955, now married, Mary and I were assigned to one of many
one-bedroom Gold Street apartments. When our son was born in 1956 we
were assigned a two-bedroom house overlooking a neighboring canyon at
4321 Fairway Drive. With our second child we were assigned to a
three-bedroom portion of a duplex house. With our 3rd child we didn't
move.

I was hired by Herman Hoerlin, renowned for his early work on cosmic
rays, as head of a newly formed section, J-10, in the Weapons Test
Division or J-division of LASL. The section was created to initiate
physics experiments treating the nuclear detonation as a high-energy
source. Ralph Speece, an electronics engineer, and I were the first
members of the section.

I soon discovered that there were two other Astrophysicists at LASL.
Ralph Williamson (PhD 1943, U.\ of Chicago) in the Weapons Design division
or T-division, and Arthur N.\ Cox (PhD 1954, U.\ of Indiana) in the Weapons
Effects section, J-15. This section was involved in studying the damage
effects from a detonation. Such things as the heat and shock load on
near and distant objects as well as nuclear radiation effects were its
purview.

During my career in J-10, I participated in four weapons test
operations. Two at the atomic bomb test site in Nevada, namely Teapot
in 1955 and Plumbbob in 1957, and one test, Redwing in 1956, at the
thermonuclear test site on Bikini Atoll in the Marshall Islands in the
Pacific, and the last one a single thermonuclear detonation, labeled
Teak, at a high altitude above Johnson Island as part of the Pacific
Hardtack Operation in 1958.

My first experiment, with Ralph, was at the 1955 Teapot series of
tests at Camp Mercury, Nevada. We planned to determine the temporal
history of air fluorescence, the so-called Teller light, induced by the
initial burst of gamma rays and later X-rays from atomic bomb devices
that were being tested. The issue was to see what diagnostics might be
learned about the nuclear reaction and to determine the emission state
of the irradiated air. The observations were performed from an
underground bunker with a surface carriage overhead that held the optics
needed to access any line of sight. The bunker was accessed by a ramp
with a blast-proof door. A quartz spectrograph with photoelectric
detectors placed at appropriate spectral locations was used.

I was excited to see the results of my first experimental effort, so
after the first `shot' I drove out to my bunker from the observing
station overlooking the Yucca Flats where the various test were sited.
There was no data! A quick check showed that the power supply to the
detectors was not turned on. So, lesson \#1, always use a Check List!
Director Hoerlin was not amused.
The subsequent results were published in a LASL technical report (Table~\ref{tab:LASLreport}.[1])

The full list of my LASL Technical Reports is provided in Table~\ref{tab:LASLreport}, and gives information to indicate the various experimental
activities with which I was involved. Some of these reports are not classified, and are available online.

\begin{table}
    \centering
    \begin{tabular}{r p{3.15in}c c}
1 & Skumanich, A. and R. Speece, 1955, 
Photoelectric Study of Teller Light and Early Bomb Light — Teapot Operation  & --- & classified \\
2 & Hoerlin, H., W. Ogle, A. Skumanich, H. S. Stewart, and R. Williamson, 1956, NRL-LASL Brightness
Temperature Experiment — Preliminary Report on Theoretical Design & J-10-188 & classified \\
3 & Skumanich, A. and A. N. Cox, 1956, On Thermal Convection in a Viscous Polytropic Atmosphere & --- & \\
4 & Cowan, R. and A. Skumanich, 1957, X-Ray Transmission and Absorption in Cold Media & J-10-284 & \\
5 & Skumanich, A., 1958, A Preliminary Analysis of the PINEX Experiment 
& J-10-359 & classified \\
6 & Skumanich, A.\ and F.\ Jahoda, 1959, Teak Experiment-Energy Deposition in Air & J-10-467 & classified \\
7 & Westervelt, D. R. and A. Skumanich, 1959, Excitation and De-Excitation in the Nitrogen Band Systems &  J-10-541 & \\
8 & Skumanich, A., 1959, Teak Fireball Formation Radiative Growth and Brightness History & J-10-566 & classified \\
9 & Skumanich, A., 1959, Radiation Pressure Acceleration of Outer Bomb Layers, \emph{Symposium on Scientific Applications of Nuclear Explosives}, Los Alamos & & classified
    \end{tabular}
    \caption{List of LASL Technical Reports (both classified and unclassified)}
    \label{tab:LASLreport}
\end{table}

One of the most unusual and complex experiments was a test of the Planck
function distribution at approximately $\simeq 10^7$ K, in both optical and X-ray regions,  filling the hohlraum using radiant energy
from a thermonuclear device at Operation Redwing at Bikini Atoll in the
Pacific (1956)

The design phase 
was led by Hoerlin (see Table~\ref{tab:LASLreport}.[2])
and included, in addition
to myself, Bill Ogle (J-Division Head), Harold S.\ Stewart (NRL-Optics
Division, a LASL subcontractor) and Ralph Williamson (T-Division).
Later, in the field execution, two new physicists hired by J-10, Don
Westervelt and Roy Blumberg along with Dennison Bancroft (PhD 1939, Harvard), a J-10 scientific consultant, contributed their talents.

One of my responsibilities was the design of the hohlraum into which the
bomb energy is dumped. The box had to be made of a high inertial mass
material, to slow down dynamic response to the rapid heating, surfaced
with a material that would vaporize due to the strong surface heating
and exert outwards pressure on the walls. Finally the box had to be
thick enough so that the propagating Marshak front (a radiation `wave', i.e., radiation field density `shock') would not break through before
the end of the measurement. The answer was a lead box with a gold plated
inner surface!

Once the box had equilibrated, the black body X-rays and visual
radiation exiting the hohlraum aperture traveled along a 1-mile-long
vacuum pipe to the detector station. Bancroft designed and supervised
the construction of the vacuum pipe system.

The experiment worked successfully and we found that the Planck function
was valid, within our experimental errors, at very high temperatures
that are unattainable in laboratory conditions.

The subsequent year (1957) at Operation Plumbbob\footnote{An eyewitness account of the activities at the Plumbbob tests can be read at \url{https://www.lrb.co.uk/the-paper/v34/n24/jeremy-bernstein/at-los-alamos}. The author of this essay, Jeremy Bernstein, was ‘inducted’ into the Los Alamos scientific ‘family’ three years after I was, and presents a good description of the 'induction' process for the interested reader.}
 in Nevada I was
responsible for another unusual experiment. I was tasked to obtain an
image of the neutron environment around the secondary of a mock-up
thermonuclear device, using a series of lead pinhole baffles, the so-called PINEX Experiment (see Table~\ref{tab:LASLreport}.[5]).

This was executed with the device on a 500-ft tall tower where I laid a
lead brick blanket with a `pinhole' aperture at various levels of the
tower to obtain an image of the neutron density at the source. The
detector to record the image was at the bottom of the tower. The
experiment was successful. The tower and lead `camera' were ultimately
vaporized and to this day I still rue the need to have so much lead
introduced into the atmosphere.

My last field experiment dealt with a single thermonuclear detonation
(labeled Teak) at a high altitude over Johnson Island as part of the 1958 
Pacific Operation Hardtack. The bomb was rocketed aloft from
Johnson Island. The Teak experiment consisted of the study of the
atmospheric physics effects of a megaton-range high-altitude energy
release.

The experiment required a variety of theoretical calculations beginning
with the X-ray dose-distance relation due to prompt radiation (see Table~\ref{tab:LASLreport}.[6]).
This provided the local excitation radiation
field as well surface level dosages. The latter indicated that to avoid
eye damage to Marshall Island natives the test had to be moved from the
Bikini test site to the isolated Johnson Island site.\footnote{See H. Hoerlin, United States High-Altitude Test
  Experiences -- A Review Emphasizing the Impact on the Environment, LASL Monograph LA-6405 (1978); available online
  at \url{https://babel.hathitrust.org/cgi/pt?id=mdp.39015086460626}.}

My task was to predict the brightness and growth of the X-ray Fireball
and to continue the study of high-energy excitation of air fluorescence (see Table~\ref{tab:LASLreport}.[8]).
The radiative evolution of the Fireball with time was calculated with
the assistance of the J-10 support scientist Fran\c{c}oise Ulam, 
wife of famed Manhattan-Project mathematician Stanislaw Ulam.

A study of possible atmospheric emission mechanisms was designed (see Table~\ref{tab:LASLreport}.[7]).
This was expedited by consultation with the LASL visiting
scientist Gerhard Dieke, the well-known molecular spectroscopist at
Johns Hopkins. His expertise in molecular spectroscopy was of invaluable
use to us because most of the luminescence was due to emission by
molecular oxygen (O\textsubscript{2}) and nitrogen (N\textsubscript{2}
and the singly ionized nitrogen molecule
N\textsubscript{2}\textsuperscript{+}) The different stopping depths of
the prompt emission and X-radiation allowed a height resolution of the
emitting molecules and hence of the atmospheric properties.

The test was successful and yielded an extensive data set. The
preliminary analysis of the air fluorescence data was published by \cite{LASL-We60}.\footnote{See also the declassified review by Hoerlin, H., Air Fluorescence Excited by High-Altitude Nuclear Explosions, LASL, Tech.\ Report LA-3417-MS (1966); available online at \url{https://babel.hathitrust.org/cgi/pt?id=coo.31924107989174}. }

With a nuclear detonation in nearly empty space the emergent radiation
field is no longer blanketed by air and one has the possibility of
radiation pressures effects at the bomb surface. I analyzed this effect
and reported on its feasibility at the 1959 LASL Symposium on Scientific
Applications of Nuclear Explosives (see Table~\ref{tab:LASLreport}.[9]).
Edward Teller, who was
advocating peaceful uses of nuclear explosives at the time, told me he found my analysis sound and of considerable interest.

A more extensive study of this effect is presented in a paper that deals
with other possible high altitude explosion effects, viz., the
interaction of radiation accelerated particles, the fireball shock
interaction with the Earth's magnetic field and generation of MHD waves,
artificial aurora, bomb plasma interactions, etc. \citep{LASL-Ar59}.

In addition to the programmatic activities at LASL there were also
others of an academic character such as frequent lectures, colloquia and
mini courses. For example, Burt Wendroff gave several talks on a new
numerical method for solving the hydrodynamic equations, Connie Longmire
gave a mini course on Plasma Physics, Nick Metropolis presented a new
approach to Monte Carlo methods while, Bengt Carlson covered numerical
solutions of the Boltzmann equation for particles including photons
(radiative transfer equation). Thus my physics understanding was
significantly extended.

Mary and I found our Los Alamos life to be culturally enriching.
Indeed, when Mary first joined me at Los Alamos, she was pleased to discover the presence of the Los Alamos Choral Society. She immediately enrolled us. One of the first pieces we studied and sang, was Mozart’s Missa Brevis in F Major (K.\,192). The cultural milieu in Santa Fe was rich with visiting Quartets and the resident Santa Fe Opera
Company.

\section{University of Rochester}

With a comprehensive test ban treaty being negotiated by President
Eisenhower in 1958 there were no 1959 or 1960 atmospheric nuclear tests.
It seemed the opportune time for a year's leave of absence. Harold
Stewart, who had left NRL to become the Director of The Institute of
Optics at the University of Rochester, urged me to come to the Institute
as a Visiting Scientist for the 1960-61 academic year. In addition,
based on Stewart's recommendation, Robert Marshak, the head of the
Physics and Astronomy department, offered me a 1-year Assistant
Professorship appointment.

In the fall of 1960 I gave my first academic undergraduate course in
Thermodynamics with a new book by H.~B.\ Callen. I had decided that it
was the best book available at the time. It presented the concept of
entropy from an entirely different point of view than that available in
other standard texts.
The book was a success with my students. I found them to be quite bright
and extremely competent and diligent. It was a pleasure to work with
them.

Coming from a nuclear weapons program it should be no surprise that my
research continued to be involved with the nuclear weapons effects
program at the Institute. A.\ Battacharjie, a co-worker at the Institute,
and I calculated, using a Monte Carlo method, the ground radiation dose
to be expected with a nuclear explosion at different heights above a
layer of clouds \cite{{josa:61}}.

Academically and culturally Rochester was a welcome interlude. It was a
pleasure to be in the depths of the academic environment. The faculty
was quite interactive so I mixed with historians, linguists,
mathematicians, and physicists. The library was a delight and as a
faculty member I had open access to the inner sanctum of the humanities,
sciences and rare books.

My wife had worked in Rochester as a registered nurse at an earlier time
which renewed previous associations. In addition she had been a member
of the Choral Society at the University's Eastman School of Music and
was able to rejoin. Furthermore, the Dryden Theater of the Eastman
Museum presented a wealth of national and international free movies. We
enjoyed the rich cultural environment. Finally, from an emotional need,
we were only hours away from our families, friends and the Rusyn culture
in Pennsylvania.

My rewarding experience at the University of Rochester led me to decide to leave Los
Alamos and weapons work and re-enter the academic environment. But here
I was in a bit of a quandary. My first opportunity was a staff position
without term at the High Altitude Observatory (HAO) of the University of Colorado, Boulder (CU Boulder)
offered to me by Walt Roberts while at Los Alamos, where he gave a
colloquium. He told me that HAO was to undergo incorporation into a new
larger institution to be dedicated to the Atmospheric Sciences in
general. This was a standing appointment.

I also explored the possibility to stay at the University of Rochester.
Marshak extended my appointment for another year. However I feared that
in such a short time and with my experience only in weapons physics I
would not be able to develop an astrophysical research program that
would ultimately lead to a permanent appointment. Perhaps I should have
argued for a three-year appointment, but I was reluctant to do so. I was
worried that even three years might not lead to a viable research
problem. So I decided to accept the position at HAO.

We (two parents and three young children) left Rochester in the late
Summer of '61, stopping to spend some time in Pennsylvania before
leaving with sad hearts for the new opportunity in Colorado.

\section{High Altitude Observatory -- Early Years}

My family and I arrived in Boulder in the first week of October 1961
after an early fall snowstorm. We were temporarily assigned a house
originally intended for an Australian visitor whose arrival was delayed.
I was duly appointed as a staff scientist at HAO and as a Lecturer in the University of Colorado Department of
Astrogeophysics on September 16, 1961. Very shortly afterwards, in
January 1962, HAO became the initial division within the National Center
for Atmospheric Research (NCAR) managed by the University Corporation for
Atmospheric Research (UCAR), and I became a staff scientist of NCAR.

Walt Roberts, the then director of HAO became President of UCAR while
John Firor was hired to the Directorship of HAO and R.\ Grant Athay became
Head of the Astrogeophysics department. John was an understanding person
and accepted my initial lack of direction for my research. Coming from a
weapons laboratory I had no ``irons in the fire'' to pursue. Even at the
University of Rochester my research was related to atomic weapon effects.

\subsection{Chromospheric Activity Decay with Age} \label{sec:CaK}

In reviewing the literature looking for research ideas, I came across a short note by \cite{1957ApJ...125..297V}
on ``Population Differences Among M
Dwarfs.'' Using the presence or absence of chromospheric \ion{Ca}{2} H-K
emission to assign class membership rather than a chemical difference,
as was done by \cite{1953AJ.....58...96V}, \cite{1957ApJ...125..297V} found similar
peculiar velocity dispersion differences. Main sequence dM
stars with H-K in emission were in the kinematically cooler, smaller
peculiar dispersions, population class, hence younger.

A previous evidence for age dependence of H-K emission dwarfs was that of
\cite{Delhaye1953}. Using local solar
neighborhood dM and dMe stars listed in the extant literature, he found
that dMe stars had a significantly smaller scale height (lower mean
velocities) perpendicular to the galactic plane than dM stars. He
surmised that ``ces \'etoiles formeraient donc un sous-syst\`eme tr\`es plat et
pourraient \^etre tr\`es jeunes.'' (these stars [...] might be very
young.) This result was treated with circumspection, as the sample
number was quite small.

This prompted me to initiate a study of the peculiar velocities of the
field dG Stars (solar like) using \ion{Ca}{2} H-K emission (presence or
absence) as the classifier, as was done by \cite{1957ApJ...125..297V}. The results indicated that
the \ion{Ca}{2} H-K emission group was younger than the non-emission group \citep{1965AJ.....70S.692S}.

Concomitant with the search for stellar chromospheric emission data for
field stars was the search for photometric data that would yield stellar
luminosity, color (a proxy for surface temperature), and degree of
chemical metallicity. I found a homogeneous photometric data set in an
unpublished Yerkes Observatory stellar catalogue by Stromgren and Perry.

In the fall of 1963 Olin Wilson gave a colloquium talk at HAO on his
investigation into \ion{Ca}{2} H-K emission intensities (eye estimates) in
Galactic Clusters and field stars. The object was his search for a
probable correlation between chromospheric activity and age in
Main-Sequence Stars. Cluster ages were derivable from their deviation
from the zero age distribution in the Hertzsprung-Russell diagram
(luminosity-color plane) but the lack of an accurate homogeneous data
set for the field stars precluded his doing so.

During the colloquium I stated that I could date his stars using the
Stromgren and Perry catalogue. One could construct an accurate and
coherent H-R diagram and thus date the emission field stars by their
location. He suggested we collaborate, and we found that indeed the
strongest emission field stars inhabited the zero age H-R region \citep{1964ApJ...140.1401W}.

Subsequently, Wilson, with a newly designed photometric detection
system, began publishing the `apparent' equivalent widths of the \ion{Ca}{2}
emission lines (widths that included the equivalent width due to the
underlying photospheric absorption line) for the Hyades cluster as well
as that of field stars.

I had found, in Olin's field star catalogue, the \ion{Ca}{2} emission data for
the Ursa Major cluster and it's associated stream stars. Luckily, an age
for this cluster was also available.

In the case of the Sun, whose age was well determined, I had Olin's
measure of the solar \ion{Ca}{2} H-K emission `apparent` equivalent width (in
angularly integrated light, i.e., reflected Moon light).

After subtracting the underlying photospheric line's equivalent width and
removing the color dependence of the normalizing continuum \citep{1981ASIC...68..349S} I had three data points, Hyades,
Ursa Major and the Sun in an emission luminosity vs.\ age diagram.

I assumed that the temporal evolution of the chromospheric emission was
in an asymptotic state and hence initial-value free, i.e., governed by a
power law. I learned of this type of behavior from my late 1950's
experience at Los Alamos. I found that an inverse square root of age
power law fitted my three points.

In 1970 Robert (Bob) Kraft was in Boulder on a sabbatical and in one of
my conversations with him I discussed my emission-age power relation
finding. Kraft suggested that his observations with Greenstein of the \ion{Ca}{2} emission equivalent widths in late-type Pleiades and Hyades would
allow one to add the Pleiades to the graph. His preliminary calculation
indicated that the Pleiades extended the square-root relation. He also
directed me to his rotation spin down data which I found also followed a
square root relation.

Finally I discovered Peter Conti's (1968) suggestion of a ''relation
between rotational braking and Lithium content'' so I added his Lithium
data to my graph.
Lithium followed a square root drop to the Hyades age but showed a
significant over-depletion for the Sun.

I wrote up my results and sent the paper to the 
Astrophysical Journal (ApJ) Letters editor, Don
Osterbrock. Don called me and urged me to withdraw my proposed letter and resubmit it as a paper to the  ApJ  main journal, and if so, he would deliver my manuscript by walking it across the hall.  With Osterbrock’s imprimatur the paper was published directly \citep{1972ApJ...171..565S}.\footnote{The
  persistent influence of this paper on Solar-Stellar Astrophysics is
  described in Sect.~\ref{sec:50th}.}  
  We note that the data reduction procedures used to arrive at calibrated data was published much later \citep{1981ASIC...68..349S}.

At the 1969 AAS meeting, Edward (Ed) T.\ Frazier presented scatter plots of simultaneous photometric observations of the solar network in a $2.4''\times 2.4''$ square region at the center of the solar disk. He plotted the relative contrast (i.e., mean-normalized brightness deviation from the mean over the map) of the \ion{Ca}{2} K core (1.1\,\AA\ bandwidth) vs the same quantity for the \ion{Fe}{1} 525.0 line. In turn, the relative contrast of \ion{Fe}{1} 525.0 line was plotted against the vertical magnetic field (inferred from \ion{Fe}{1} 523.3). From my seat at the meeting, my ‘eyeball’ fit to the scatter plots implied a linear relation between \ion{Ca}{2} K contrast and the magnetic field.

This sparked my interest in deriving a direct relation between \ion{Ca}{2} K contrast and the magnetic field. I knew that I could calibrate the \ion{Ca}{2} K relative contrast to derive the emission equivalent width index and examine its dependence on the magnetic field.

I proposed to Ed that I do so and he agreed to share his data. We found, after a detailed analysis, that, indeed, the \ion{Ca}{2} K
emission index was linearly dependent on the magnetic field \citep{1975ApJ...200..747S}. 

This conclusion coupled with the Skumanich square-root law \citep{1972ApJ...171..565S} led to interesting implications regarding magnetic braking and the spin-down of stars similar to the Sun.
This paper was completed during my stay at the Laboratoire de Physique
Stellaire et Planétaire (LPSP) of the Centre National de la Recherché
Scientifique (CNRS).

\subsection{Radiative Transfer and Resonance Spectral Lines}
\label{sec:RT}

The formation of resonance spectral lines, i.e., the prediction of their
shape and strength in the emergent radiant spectrum of the Sun, depends
on the equation of transfer (the Boltzmann evolution equation) that
allows one to calculate the radiation field (the radiative Boltzmann
function) through the Sun's outer layers given the emission and
absorption coefficients for the spectral line in question. These
coefficients depend on the number of atoms that occupy the internal
energy states that define the energy of the line. They are determined by
population rate equations that contain the detailed processes,
collisional and radiative, that populate or depopulate the energy state.

In the case of a steady state, detailed balance holds and the population
rate must equal the depopulation rate so that with conservation of the
total atom density one has a single equation for the population ratio of
the two energy states involved.

The radiative process in the population equations is, in terms of
kinetic theory, given by the binary product of the local photon density
and exciting particle and the interaction (absorption) cross-section.
The photon density is defined as the angular integral of the Boltzmann
photon function or its equivalent, the specific intensity. The latter
usage leads to the term `\emph{mean} specific intensity'. I find it
preferable to use the directly understandable photon density
designation.

Following the Harvard astrophysical approach one represents the photon
density by an integral operator, the so-called `Lambda' operator, which
maps the source function into the monochromatic photon density. This
leads to a non-linear system of equations whose solution yields the
ratio of occupation numbers (i.e., emission to absorption coefficients)
--the so-called source function.

I received a letter from Grant Athay dated 19 Aug 1965 where he writes:
\emph{``I stand in the peculiar and unenviable position of claiming that all
existing theories of line formation, including all solutions for the
two-level atom, are devoid of physical meaning. I now have one convert
to my cause, viz.\ [J. C.] Pecker. If you are willing [...] meet me in Munich.
We (you and I) have some exciting work to do.''}
This was a remarkable assertion and stirred my interest. 

I soon met
Grant in Munich and he went through his argument with me. I began
studying his set of equations. Ultimately I found errors in his analysis
that when corrected negated his conclusion. The world was not to be
changed.

At one point in his argument Grant made use of the gradient of the
radiant flux-density that occurs in the first angular moment of the
transfer equation, i.e., an energy conservation equation.

On reflection it occurred to me that one could replace the photon
density in the population equation with its form as defined by the
energy equation rather than by its primitive definition. This introduces
the gradient of the flux-density into the population equation as an
integro-differential operator, refer to \cite{1963bmtp.book.....K}, acting on the
source function. Because its kernel has shorter range than that for the
Lambda operator one has better convergence properties.

\begin{figure}
\centering
\includegraphics[width=.99\textwidth]{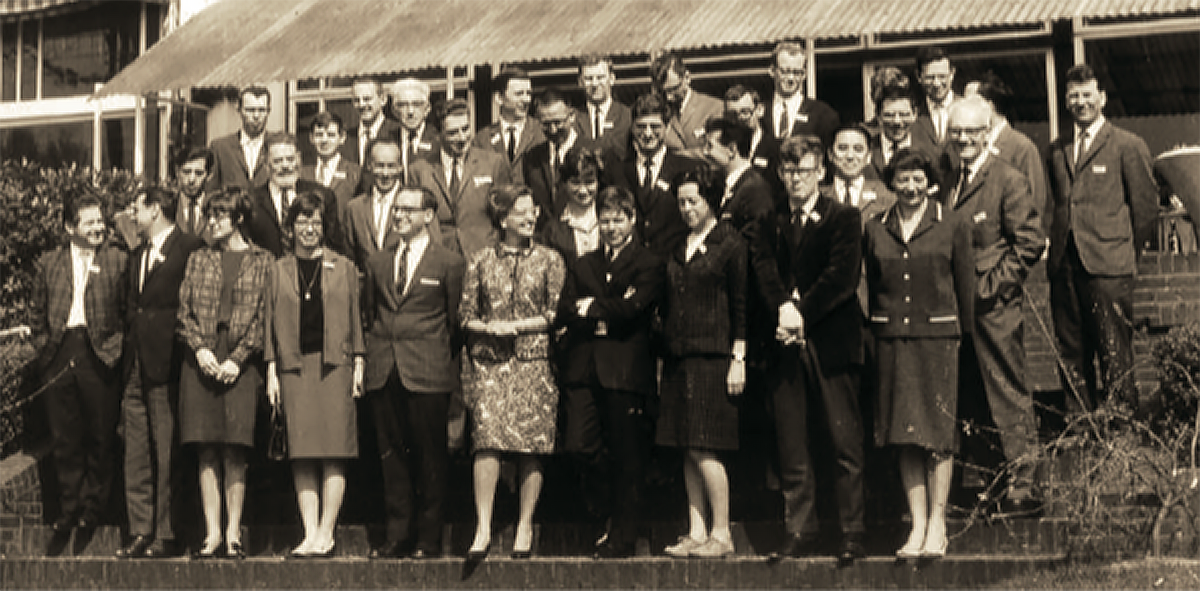}
\caption{Participants at the Bilderberg conference, April 1967.
Front row, left to right: Jean-Claude Pecker, 
John T.\ Jefferies, Carla E.\ Boot (support staff), Marijke Burger, Robert W.\ Noyes, Edith A.\ M\"uller, 
Roger M.\ Bonnet, Simone Dumont, David L.\ Lambert, Yvette Cuny. Second row: Jacques E.\ Blamont, Jacob Houtgast, R.\ Grant Athay, Christiane Guillaume, Nicolas Grevesse, Osamu Namba, Cornelis de Jager.
Third row: J.\ Paul Mutschlecner, Owen Gingerich, George Withbroe, Marcel G.~J.\ Minnaert, Tom de Groot, Hartmut Holweger, Jacques Sauval, Hans Vesters (support staff), Pierre Souffrin, Hans Hubenet, Michel Hers\'e, Robert J.\ Rutten, Pierre J.\ L\'ena, 
Andrew Skumanich, Philippe Delache, Jaap B.\ Vogel (support staff).}
\label{fig:Bildberg}
\end{figure}

At Los Alamos I found that inversions based on the energy equation were
more stable than using the primitive integral representation of the
photon density. So, I proposed to Grant that we use my formulation given
above to obtain better source functions (i.e., population) solutions.
The analysis and derivation of the numerical algorithm is given by \cite{1967AnAp...30..669A}\footnote{The following author attribution
  is listed therein. ``In this and in subsequent papers we adopt the
  convention of listing the authors names alphabetically without
  implications as to senior authorship.''} and was programmed for the
Cray by William B.\ (Buck) Frye, a HAO research assistant.

With a robust and accurate flux-gradient inversion algorithm, Athay and
I undertook to find a thermal (kinetic) model that reproduces the
observed \ion{Ca}{2} emission features. We were successful and the results
appeared in the series of papers \citep{1968ApJ...152..141A,1968SoPh....3..181A,1968SoPh....4..176A,1968AJS....73S...2A,1968ApJ...152..211A,1969ApJ...155..273A}. How
accurately these models represent reality depends on how good the two
energy state model of the singly ionized calcium captures reality.

Our results were presented to the 1967 Bilderberg Conference (see Fig.~\ref{fig:Bildberg}) and helped
to define the nature of the low chromosphere. At this conference I met
Roger Bonnet, director of Laboratoire de Physique Stellaire et Planetaire
(LPSP), as well as Yvette Cuny, on the scientific staff of Service
d'Astrophysique de l'Observatoire de Meudon. Roger presented his space
observations of the solar UV spectrum. Yvette reported on her successful
determination of Non-Thermodynamic Equilibrium structure of a hydrogen
model of the solar atmosphere. Both became my lifelong colleagues and
friends.

\begin{figure}
    \centering
\includegraphics[width=.98\textwidth]{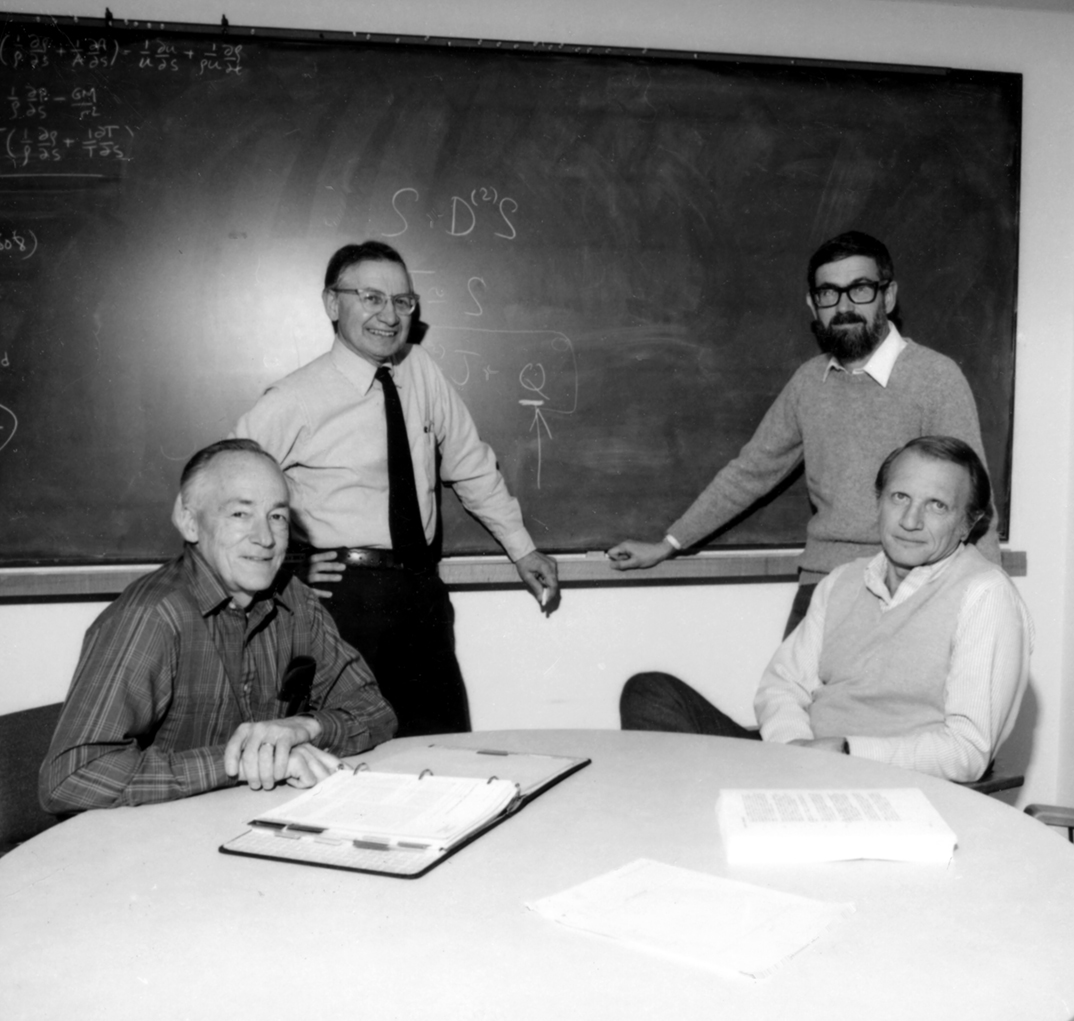}
    \caption{Members of the HAO Radiative Transfer Applications Section (c. 1973). From the left, Grant Athay, Andrew Skumanich, Dimitri Mihalas (standing), Lew House (sitting).}
    \label{fig:best-of-HAO}
\end{figure}

The theoretical investigation of radiative transfer became a dominant subject at HAO - and has remained as such to these days - and led to the formation of a dedicated group of scientists, which included, in addition to myself, Grant Athay, Lew House, and Dimitri Mihalas (Figure~\ref{fig:best-of-HAO}).

\subsection{Orbiting Solar Observatory OSO-8 (1975--1978)} \label{sec:OSO-8}

During my 1973 visiting scientist position at LPSP I was invited by
Bonnet to participate as a co-investigator in the French UV
spectroscopic experiment on the NASA orbiting solar satellite OSO-8
launched in June 1975. It carried in its pointed section a LPSP
multichannel spectrometer designed for the highest angular and spectral
resolution achieved by spacecraft at the time. The chromospheric
resonance lines of \ion{Ca}{2}, \ion{Mg}{2}  and \ion{H}{1} as well as the transition region
lines of \ion{O}{6} and \ion{Si}{3} were to be observed.

The observing team was composed of R. M. Bonnet, P. Lemaire, J. C. Vial,
G. Artzner, P. Gouttebroze, A. Jouchoux, J. W. Leibacher, A. Skumanich,
and A. Vidal-Madjar. All are from LPSP except for Leibacher (Lockheed
Palo Alto Research Laboratory, Palo Alto, CA.) and A. Skumanich (High
Altitude Observatory, National Center for Atmospheric Research and
University of Colorado, Boulder, CO).

The pointed section of the satellite operated in real time from an
observing center housed at the Laboratory for Atmospheric and Space
Physics (LASP) of the University of Colorado. LASP had the other UV
spectroscopic experiment in the pointed section. The Boulder center had
direct real time access to the dedicated services of the OSO-8 Control
Center at the Goddard Space Flight Center.

The control of the LPSP pointing system rotated among the team members.
The need for continuous observation of the Sun was an exhausting
experience. Indeed, in July 1975, upon my return from a scientific visit
to the Observatoire de Nice, I was immediately plunged into OSO-8 operations
where I was submerged for the entire year. To make matters worse,
through an oversight, I had committed myself to teach a first-year
graduate course in Modern Astrophysics, listed by the Departments of
Physics and Astrophysics, Aerospace Engineering and Astrogeophysics, for
the spring of 1976. The upshot was that I was working seven days a week
during the spring. However, I had the honor of winning the prize bottle
of Bourbon for having the most frequent successes in acquiring targets
on the first try.

In November 1975, a joint Sunspot Observing program was implemented by
Skumanich (chromosphere, LPSP-CNRS), Bruce Lites (transition region, CU Boulder)
along with Lew House and Tom Baur (HAO-NCAR) who were to provide
magnetic fields derived from concurrent ground-based Stokes polarization
observations. My pointing talents were `spot-on' and we harvested a
wealth of data for a large inactive spot with a homogeneous umbra.

The OSO-8 spot observations of \ion{H}{1}, \ion{Ca}{2} and (previously unobserved) \ion{Mg}{2} yielded a unique upper chromosphere and transition region thermal
model similar to that of the quiet-Sun, which implied that the intense
magnetic field played only a passive role for the chromospheric heating
mechanism \citep{1982ApJS...49..293L}. The thermodynamic model
derived from the UV data was unique and led to many
citations \cite[see, for example,][]{2010MmSAI..81..604K}.

The updating of the thermal parameters in this case and future others
was made significantly simpler than that in general use at the time by
the application of a sensitivity analysis procedure to the multi-line
equations \citep{1986ApJ...310..419S}. This allowed one to
determine the most important interlocking radiative transition for the
line in question that needed to be changed.

My interest was also captured by the serendipitous observation of 
simple two ribbon class-C flare \citep{1977BAAS....9..432J}. After
collecting radio, X-ray, and hydrogen H$\alpha$ data I presented an
initial analysis at the OSO-8 workshop \citep{1977OSO8}.

A review of the performance and preliminary OSO-8 results is presented
by \cite{1978ApJ...221.1032B}. This was a soft-publication option to
get data out to the community as soon as possible.

\section{High Altitude Observatory -- Later Years}

\subsection{Magnetic Regions and Stokes Analysis}

In early 1981 Athay, as acting head of the Stokes - Magnetic Fields
Program, asked me to take on the scientific responsibility for the
Disk Fields Analysis Section. The responsible program scientist had left
and the extant magnetic field inversion algorithm, called AHH, failed in
the inversion more often than not. My first challenge would be to update
or replace the AHH Stokes-inversion algorithm. The subsequent task would
be to analyze part of the archived Stokes data.
I thought this to be a challenge and decided to drop my OSO-8
involvement. This plunged me into the world of Stokes Polarimetry.

My entry into the field was facilitated by HAO visiting scientists at
the time, David E. Rees (
Department of Applied
Mathematics of the University of Sydney, Australia) and Egidio Landi
Degl'Innocenti (University of Florence, Italy). They became colleagues
and friends.

The incorporation of polarization or photon spin to describe the
radiative state introduces a 3 dimensional quantity, (the two amplitudes
of the two circular spin states and their relative phase difference), which with
the scalar specific intensity leads to a 4-vector radiative Boltzmann
`function' (the Stokes 4-vector) which satisfies a vector transfer
equation with the scalar absorption coefficient replaced by a generalized
`absorption' matrix.

I note that the early 1980's versions of the vector transfer equation and the `absorption' matrix 
were plagued by various sign errors and other associated handedness issues for the circular states. A complex quantum theory formulation by \cite{1983SoPh...85....3L}  resolved these issues and provided a physically consistent `absorption' matrix, called the Propagation Matrix.

Later, \cite{1989ApJ...343..920J}, here after JLS, derived the propagation matrix and vector transfer equation
using classical EM theory, which allowed one  to understand the physical origin of the various terms in the quantum-mechanical treatment.
Their results yielded an expression for the propagation matrix, which proved to be in agreement (after introducing quantum oscillator strengths) with the results of \cite{1983SoPh...85....3L}.

The physics that JLS used in their construction of the Propagation Matrix entails the fact that the interaction of the EM wave with the magnetic field is mediated by the complex index of refraction for each polarization component. This index appears only in the spatial term of the phase function. The imaginary part leads to extinction while the real part represents the magneto-optical effects.

Thus one only needs to increment the phase at the current spatial location $z$ along the ray, and use it as the phase at the new point, $z+dz$. A factoring  of the exponential with this new phase yields the product of the exponential of the original phase with the exponential of the differential phase. This second exponential function is linearized, which yields a multiplicative factor of the vector amplitude at $z$ to obtain the vector amplitude at $z+dz$. The linear part of the multiplier contains the index of refraction (see JLS, eq.~(3)). The difference of the amplitudes ultimately leads to the Propagation Matrix. 

\subsection{A New Stokes Inversion Algorithm}

The vector transfer equation is easily integrated (see JLS, eq. (39))  to obtain analytic Stokes
profiles (primitive profiles) for a simplified thermal structure,
represented by a linear source function vs.\ optical depth, permeated by a
uniform magnetic field, with a propagation matrix with constant atomic line
parameters. The two source function parameters, the vector magnetic
field, and the absorption shape parameters are then determined by a least-square
fit of the analytic profiles to the observed Stokes profiles.

The HAO authors of the problematic AHH inversion scheme \citep{1977SoPh...55...47A} made a number of simplifications that reduced the number of
free parameters to be fitted. They transformed the line intensity to an
absorption line depth, normalized by the depth at line center, which
eliminated the two source function parameters but at the expense of
introducing a greater degree of non-linearity. This restricted their fit
to the three Stokes variables, normalized line depth, linear and
circular polarizations, as a function of wavelength.

Thus a new algorithm was required that used the primitive equations with
all four Stokes variables and introduced more physics into the propagation matrix, namely a line profile with damping and
magneto-optic birefringence. The specific intensity would be fitted to
determine the source function parameters as well as any parasitic light
or zero point.

My development of a new HAO algorithm was joined by Rees and Landi Degl’Innocenti. Athay also asked Bruce Lites, who was working at the Sacramento Peak Observatory (SPO), in Sunspot, NM, and was a member of the HAO-SPO Stokes Consortium, to join in the development.

Rees was instrumental in my being awarded a semester-long Visiting
Scientist position in the Department of Applied Mathematics, University of Sydney in
1983. There I met Graham Murphy who was a graduate student being
supervised by Rees. His interest was in Stokes polarimetry. We
soon developed a friendship as well as a mentor-student relationship.
For Mary and me, the stay in Sydney was enjoyable. We had access to the famous Sydney Opera house as well as great Fish \& Chips bars.

The new HAO algorithm (hereafter, SL84) was tested by
\cite{1985NASCP2374..341S} on 1975 Stokes-II observations of a
large OSO-8 spot (see Sect.~\ref{sec:OSO-8})  previously inverted with AHH. It was
found that the added line opacity parameters yielded more plausible line
strength values than those derived by AHH.

\cite{1985NASCP2374..306S} further tested the SL84
inversion on synthetic Stokes profiles for \ion{Fe}{1} $\lambda 5303$ calculated with
an extant physical model of a sunspot with an assigned vector magnetic
field. The inversion field parameters successfully represented the
vector magnetic field to within a few percent, except that the derived
line opacity parameters differed considerably and were a poor diagnostic
of the thermodynamic state. However the derived source function slope
was representative of that in the spot model.

Much later, \cite{1987ApJ...322..483S} published an updated
version of the SL84 algorithm along with an extended comparison of the
OSO-8 spot inversions, with a successive introduction of each new SL84
parameter, with the AHH inversion. It was found, for example, that the
AHH magnetic field was 30\% below the SL84 value derived with the full
SL84 parameter set.

Thus we found that the SL84 inversion method was robust and that it
provided more reliable and accurate estimates of sunspot vector magnetic
fields without significant loss of economy than any other extant
methods.

\subsection{Stokes-I and Stokes-II Science}

Our first effort (early 1986) was to analyze the diagnostic usefulness
of the neutral \ion{Mg}{1} $\lambda 4571$ line using Stokes-I 1978 data. This was
initiated as a collaborative effort with Rees and Murphy, who were in
Australia at the time, where we explored the magnetic and thermal
information content of the $\lambda 4571$ line. This line is formed in the
relatively unexplored umbral temperature minimum region of sunspots. The
line is meta-stable and is collisionally dominated, and can be represented
by a linear source function.

In \cite{1987ApJ...318..930L} we found that the temperature minima
of umbrae were more extended in height than previously believed. Hence
the spot's chromosphere was optically thinner than extant models
indicated. In addition, while the Stokes profile of $\lambda 4571$ is a sensitive
diagnostic of the minimum temperature, it is also very sensitive to the
assumed value of the absorption line wing-damping constant, which is not
a well-known atomic parameter for this line. For this reason, the $\lambda 4571$
line was considered of limited utility as a diagnostic of magnetic
fields.

In the fall of 1986, I was reappointed a Visiting Scientist in the
Department of Applied Mathematics of the University of Sydney while Bruce was awarded
a Visiting Scientist position. While there we initiated a new
collaborative study with Rees and Murphy to invert the Stokes-I sunspot
umbral data of the strong Magnesium 
\ion{Mg}{1} b lines at $\lambda 5172.7$\,\AA\ and
$\lambda 5183.6$\,\AA. The analysis was finished at HAO with Rees as a visiting
scientist and Murphy as a Newkirk Graduate Research Assistant.

The \ion{Mg}{1} b lines are sufficiently strong that their source function
cannot be approximated by a linear function of optical depth. An
exponential form had to be added to the linear term to approximate the
source function predicted by non-LTE computations for an extant model of
a sunspot umbral atmosphere. The exponential was used because it was
integrable and adds a new analytic component to the existing SL84 least
squares inversion scheme.

The new inversion methodology was tested on synthetic Stokes profiles
derived, as before, for a realistic sunspot thermal model. We found that
the new scheme was most effective in recovering the magnetic field used
in the synthetic spot model Stokes profiles if one restricts the fit to
the Doppler cores of such chromospheric lines and fits both the
intensity profile as well as the polarization profiles. In addition it
was necessary to fit both members of the multiplet with their different
Zeeman splitting patterns in order to obtain the optimal comparison with
the synthetic field.

Applying the revised inversion scheme to the 1978 Stokes-I data \citep{1988ApJ...330..493L}, one found reasonable values for the magnetic
field but only if an additional ``macroturbulent'' profile smearing was
introduced, as well a correction of the observed intensity profiles for
stray unpolarized light. Due to the Stokes-I polarimeter limited spatial
resolution the results were considered descriptive rather than
definitive.

The value of this multiplet is that it is the most favorable of pairs of
chromospheric lines for quantitative analysis of chromospheric magnetic
fields.

Note that the analysis methodologies introduced in the
magnesium papers \citep{1987ApJ...318..930L,1988ApJ...330..493L} have been recently adopted and
extended to magneto-hydrodynamic simulations by \cite{2022A&A...659A.156D}.

The ultimate analysis of the archival Stokes data was the inversion of
the 1980 Stokes-II observations of four large sunspots. The data
contained only the $\lambda 630.26$ nm line of the \ion{Fe}{1} multiplet.

\cite{1990ApJ...348..747L} found that the magnetic field
occupied significant fractions of the area with both light and dark
penumbral filaments, and that the intrinsic field has a similar
threshold value both in the sunspot penumbra and in the surrounding
plage areas.

The variation of the poloidal field strength and inclination with
distance from the center of the symmetric sunspots exhibited little
non-potential character and was well represented by the potential field
of a buried dipole except in the case of spots with twisted fields.

This work demonstrated that it was possible to invert the Stokes
profiles of a single solar absorption line and to derive the
corresponding vector magnetic fields with considerable confidence in the
results. This optimistic situation was welcome from the standpoint of
advancement of our under-standing of MHD processes acting in the solar
atmosphere, and as such it further validated efforts to improve the
resolution of the Stokes observations.

\section{ASP Science}

\subsection{Advanced Stokes Polarimeter}

The successful vector magnetic field results from the Stokes-II data
inspired the drive for a new polarimeter, to be called the Advanced
Stokes Polarimeter or ASP. For this purpose, an ASP program was
established as a joint project between the National Solar Observatory
(NSO) and the High Altitude Observatory (HAO). It became a part of a
larger thrust for Stokes polarimetry being advanced on instrumental,
analysis, and theoretical fronts by a new Stokes Consortium, namely, the
HAO, NSO, University of Hawaii, and the Astrophysical Observatory of
Arcetri, Florence, Italy. Lites was appointed Instrument Scientist to
lead the HAO project. I continued my theoretical role.

The ASP would have updated components and would use the
SPO Vacuum Tower Telescope (VTT) and SPO spectrograph to
obtain higher spatial, spectral and temporal resolution than was
possible with the Stokes-II instrument. The use of CCD
cameras would allow one to obtain simultaneous Stokes
parameters of the solar image along the spectrograph slit rather
than at a single point as in Stokes-II. To build up a Stokes map
one would move the solar image across the slit by slewing the
telescope.

In addition to the polarimeter itself a Telescope Model would need to be
devised to compensate for the spurious polarization introduced by the
VTT. Such a Telescope Mueller matrix, whose inverse is needed, proved to
be fairly complex. The calibration model is described by \cite{1997ApJS..110..357S}.\footnote{This model was also used by NSO at the inception of the design of the DKIST telescope
  (c. 2005).} The final polarimeter instrument is detailed in \cite{1992SPIE.1746...22E}.
The ASP system was commissioned in 1992 and became available to the
scientific public.

The HAO effort to develop a complicated and expensive ASP program was
not without its detractors. At an AAS meeting Harold Zirin (Big Bear
Observatory) approached me (standing at my ASP poster display) to
criticize HAO for wasting federal money on a project that was not really
needed. He said his work with a filter-based magnetograph in
conjunction with H$\alpha$ images produced magnetic data whose functional
use was superior to that from ASP.

\cite{1994SoPh..155....1L} compared the
inversion results from the ASP data with those derived from a
filter-degraded version of the same data. Significant differences were
found particularly more so in areas where `filling factors', which are not
accounted for in filter magnetographs, are important. \cite{1996SoPh..163..223L} gives a more direct response to Zirin's
criticism.

Initially Bruce and I were the only members of the ASP Analysis Group.
Valentin Martínez Pillet, a HAO visitor on leave from the Instituto de
Astrofísica de Canarias (La Laguna, Tenerife, Spain), joined us for an
extended period of time. K.~D.\ Leka, a 1994--1997 Postdoctoral Fellow
in the NCAR Advanced Study Program, also joined the group.

We had help from Paul Seagraves, the HAO Senior Programmer,
on software development. Other visitors or users often joined us as
collaborators or conversely we became their collaborators.

Bruce, as Instrument Scientist, was responsible for the public use of
the ASP facility and had help from David Elmore (HAO Senior Engineer) on
facility glitches.

The ASP became the most subscribed instrument at the Dunn Solar Telescope. The instrument concept was put on the Japanese-US Hinode Satellite with Bruce Lites as the PI. The validation of precision spectro-polarimetry with the ASP provided essential scientific arguments for the justification to build a 4-meter solar telescope (DKIST).

\subsection{Magnetic Field Expulsion} \label{sec:MFE}

The most singular event observed with the ASP was in 1992 during the
passage across the disk of a decaying active region, NOAA 7201. After
the disappearance of the trailing spot one observed in the same area the
emergence and rise of a nearly closed magnetic system \citep{1995ApJ...446..877L}. The intersection of the system
with the photosphere yielded a topographic sequence from which the 3D
structure of the magnetic field was found.

The initial flux emergence region shows a rather simple geometry, but it
subsequently develops a small $\delta$-sunspot configuration\footnote{A spot
  that contains umbrae of opposite magnetic polarity within a single
  penumbra.} with a highly sheared vector field along the polarity
inversion line running through it. The magnetic system persists in the
corona well after the dark $\delta$-sunspot has disappeared from the
photosphere.

Observations of associated H$\alpha$ prominences \citep{1995ApJ...446..877L} and accompanying X-ray emission by Yohkoh \citep{1992PASJ...44L..63T} of the event
indicated a similarity to a Coronal Mass Ejection (CME).

B.~C.\ Low (HAO) found that the magnetic field could be described by an
analytic three-dimensional magnetostatic model representing a closed,
spheroidal magnetic system, in which the Lorentz force arising from
cross-field currents is balanced by the gravitational and pressure
forces.

At this point in the history of this event, BC had a flash of genius and
realized one could develop a time-dependent evolution by invoking the
concept of self-similarity. He explored this approach with the HAO
post-doc Sarah Gibson and the consequence was a new model of CMEs. The use of this model to analyze individual CMEs
transformed the nature of Space Weather research.

\begin{table}[t!]
\caption{ASP observational highlights}
\label{tab:obshigh}
\begin{tabular}{lll}

\hline
\textbf{Project} &
\textbf{Deductions} &
\textbf{Publications} \\
\hline

\begin{minipage}[t]{0.29\textwidth}\raggedright
Observations of the mesoscale magnetic structure of sunspots
\end{minipage} &
\begin{minipage}[t]{0.35\textwidth}\raggedright
Self-similar fields, size invariance (e.g., penumbral-field inclination
invariance)
\end{minipage} &
\begin{minipage}[t]{0.25\textwidth}\raggedright
\cite{1992ASIC..375..121S}
\end{minipage} \\

\hline

\begin{minipage}[t]{0.29\textwidth}\raggedright
Fine scale structure of a sunspot
\newline

Optical tomography and magnetic structure
\end{minipage} &
\begin{minipage}[t]{0.35\textwidth}\raggedright
Bright and dark penumbral `spines'
\newline

Spine elevations azimuthally corrugated
\end{minipage} &
\begin{minipage}[t]{0.25\textwidth}\raggedright
\cite{1993ApJ...418..928L}
\newline\newline

\cite{2001ASPC..236..543S}
\end{minipage} \\

\hline

\begin{minipage}[t]{0.29\textwidth}\raggedright
Downward mass flux in the penumbra
\newline

Optical tomography and magnetic structure
\end{minipage} &
\begin{minipage}[t]{0.35\textwidth}\raggedright
Arched penumbral spines with downflow
\newline

Lower Layers with Return Flux
\end{minipage} &
\begin{minipage}[t]{0.25\textwidth}\raggedright
\cite{1997Natur.389...47W}
\newline

\cite{2001ASPC..236..543S}
\end{minipage} \\

\hline

\begin{minipage}[t]{0.29\textwidth}\raggedright
Properties of magnetic flux at the site of emergence
\end{minipage} &
\begin{minipage}[t]{0.35\textwidth}\raggedright
Pore formation 
\end{minipage} &
\begin{minipage}[t]{0.25\textwidth}\raggedright
\cite{1998AAp...333.1053L}
\end{minipage} \\

\hline

\begin{minipage}[t]{0.29\textwidth}\raggedright
The evolution of pores
\end{minipage} &
\begin{minipage}[t]{0.35\textwidth}\raggedright
Pore to small spot transition 
\end{minipage} &
\begin{minipage}[t]{0.25\textwidth}\raggedright
\cite{1998ApJ...507..454L}
\end{minipage} \\

\hline

\begin{minipage}[t]{0.29\textwidth}\raggedright
The evolution of magnetic structures in terms of size-flux relationship
\end{minipage} &
\begin{minipage}[t]{0.35\textwidth}\raggedright
Pore flux-size scaling 
\end{minipage} &
\begin{minipage}[t]{0.25\textwidth}\raggedright
\cite{1999SoPh..188....3L}
\end{minipage} \\

\hline

\end{tabular}
\end{table}

\subsection{Spots And Pores}

Table~\ref{tab:obshigh} lists the observational highlights of the ASP related to spots and pores.

The ASP data demonstrated that penumbra
have fine scale structures and associated dynamical behaviors quite
different from the extant understanding. The penumbras were found to
contain narrow radial `spines' of more intense magnetic field more
vertically oriented than their surrounding field. This second field,
more horizontal, flux tube component with its associated strong downward
material motion, arches downwards into the photosphere at the edge of
the spot.

In the case of pores it was found that their size was related to their
magnetic flux content and that the field inclination at the pore
boundary increases as total flux increases. Both of these observations
support the recent pore models proposed by \cite{2000MNRAS.314..793H}.

The unprecedented ASP results caught the attention of an international
group of mathematicians, Nigel O.\ Weiss (University of Cambridge), John
H.\ Thomas (University of Rochester), Nicholas H.\ Brummell (University of
Colorado), and Steven M.\ Tobias (University of Leeds).
They argued \citep{2004ApJ...600.1073W} that the field
lines that plunge below the solar surface near the edge of the spot are
pumped downward by small-scale granular convection outside the sunspot.
They also found that such magnetic pumping could explain the abrupt
appearance of a penumbra for pores above a certain flux level.

Note that \cite{2004ApJ...600.1073W} gave an excellent synthesis of the
relevant ASP observations in addition to the contributions listed in
Table~\ref{tab:obshigh}.

\subsection{Observing the Previously Unobserved} \label{sec:unobserved}

Two new magnetic manifestation were discovered in ASP observations,
namely Horizontal ``Internetwork'' Fields \citep{1996ApJ...460.1019L} and Plage Azimuth Centers \citep{1997ApJ...474..810M}.

Due to the high linear polarization sensitivity of the ASP, one was able
to detect weak transverse fields. This allowed us to discover that quiet
regions near the center of the solar disk were found to contain
transient small-scale (typically 1"-2" or smaller), predominantly
horizontal magnetic flux features that often occur between regions of
opposite polarity (but weak) Stokes circular polarization profiles.
These features occur in isolation of the well-known, nearly vertical,
flux concentrations usually seen in the photospheric ``network.'' Hence
these small-scale horizontal ``internetwork'' fields were labeled as
HIFs.

We view the HIFs as the emergence of loops of flux, carried upward
either by granular convection or magnetic buoyancy. Even though these
entities show weak field strengths, they also seem to be fairly common
and may contribute significant flux to the upper atmosphere.

The ASP has, for the first time, allowed one to observe the vector
magnetic field of active region plages\footnote{Plage fields are
  believed to represent an average over a collection of elemental flux
  tubes (as yet unobserved) in the resolution window.} with an angular
resolution of approximately 1 arcsecond. It was found that maps of the
azimuth of the regions vector field shows magnetic structures that are
small-scale, several arcseconds, that have the full 360-degree azimuthal
distribution corresponding to diverging (or converging) field lines from
a magnetic center similar to spots.

Some such structures are associated with darkening in the continuum and
are examples of normal pores. In many cases these structures do not show
darkening in the continuum. We designate such features ``azimuth
centers'' (ACs). Such structures could only be revealed by the type of
precise measurements of the linear polarization obtainable with the ASP.

The AC structure appears to represent an intermediate state between
elemental flux tubes and pores (or small sunspots). Within ACs, the flux
tubes presumably still retain their identity, but the magnetic flux
tubes have been concentrated together, perhaps by a converging flow that
imposes a magnetic center. The magnetic flux in an AC is not large
enough to strongly inhibit heat transfer to the upper layers, so the
structure has normal photospheric brightness. Their role in the magnetic
evolution of solar magnetism remains to be explored.

\section{Solar-Stellar Chromospheres}

\subsection{Solar Activity}

With the advent of the International Ultraviolet Explorer (IUE), circa
1980, a large data set of UV irradiances (radiant flux-density at the
observer) for Solar-like stars became available. Further more, episodic
measurement of certain solar UV irradiances, which play an important
role in terrestrial atmospheric studies, were also available.

Hence there was a need for a general model to calculate any such
irradiances for the Sun where surface sources could be uniquely
identified and whose radiance (specific intensity) could be determined.

The first attempt to statistically identify such sources and determine
their absolute chromospheric \ion{Ca}{2} K emission radiance appeared in \citealt{1975ApJ...200..747S}.\footnote{This paper has been last
  cited in 2022. It is the second most cited paper in my bibliography
  after my spin-down paper \citep{1972ApJ...171..565S}.} From a study of a
magnetically quiet region at the center of the `disk' it was found that
two sources, the super-granulation `cell', which is essentially free of
elemental magnetic flux tubes, and `network', a loose collection of
elemental flux tubes at the downflow lanes of the super-granulation,
could be uniquely identified and their absolute radiance determined.

To obtain any particular activity state in the solar activity cycle one
must add the contribution of plages as a 3rd component, to the minimum
state. This requires knowledge of the radiance as well as the size and
positional property of each plage. Fortunately for the former, the
calibrated synoptic observations by Dick White (HAO) and Bill Livingston
(KPNO) were available. For the latter, the WDC/NOAA \ion{Ca}{2} K plage
records sufficed. These were provided by Judith Lean, a post-Doc at the
Cooperative Institute for Research in Environmental Sciences (CIRES),
University of Colorado, Boulder, because of her interest in UV
terrestrial irradiances. All three joined me in this effort.

It was found that a successful agreement between the constructed K
irradiance and the measured irradiance was possible only if the
(elemental) flux dispersed from the breakup of plages was added as a
third `active network' component. The full description of the new K
irradiance model is presented in \citealt{1984ApJ...282..776S}.

A well-received review of the long-term nature of solar variability
circa 1981 can be found in \cite{1981ASIC...68..349S}.

\subsection{A unique Ca~II K emission Dwarf M Star (dMe)}

In 1982 Arthur Young, Professor of Astronomy at San Diego State
University, came to HAO on a sabbatical leave. After a number of
discussions with him regarding his solar-stellar connections research he
invited me to join him in the analysis of his recent observations of
both \ion{Ca}{2} K and H$\alpha$ emission in dwarf M (dM) field stars. Our initial
effort indicated anomalous properties of a number of individual stars \citep{1984ApJ...282..683Y}. Discovered here was Gliese
890, classified as dM2.5e (`e' for the presence of K emission) and
estimated to be a fast rotator.

To fully investigate GLS 890 we initiated a series of observations using
the 2.1 m telescope, Coude spectrograph, and CCD detector at the Kitt
Peak National Observatory. This was a pleasurably experience for me as
my last observing experience was with Olin Wilson back in 1964.

The analysis of our various observations
\citep{1984LNP...193..112Y,1986LNP...254..127Y,1990ApJ...349..608Y}
indicated a single star with the shortest
rotation period known which had a photometric wave due to sunspots and a
non-uniform distribution of H$\alpha$ emission plages on its surface. The
plage and spot activity appear to be at high latitudes compared to the
occurrence of active regions on the Sun. Whether the extremely rapid
angular velocity (60 times solar) confines such activity to the polar
latitudes, or merely permits it to migrate to such latitudes, could not be established
through our observations, which were confined to just the high latitude phase of the activity cycle.
Perhaps we are observing the activity of a different type of dynamo than
the solar one.

\subsection{dMe Stars in General}

My last venture in stellar activity with Young was to investigate the
energetics of the most active population of field dMe stars. Extant
evidence from observations conducted at the McCormick Observatory \citep{1957ApJ...125..297V} indicated that
of the general population of dM stars ($N=305$) in the solar
neighborhood 21\% were \ion{Ca}{2} K emission stars, or dMe stars. Of those,
20\% also had the H$\alpha$ line in emission. The kinematic evidence
indicated that this third group was younger than the dMe stars.

Our investigation of this third state \citep{1989ApJ...344..427Y}
used H$\alpha$ equivalent-width data from three observatories \citep{1986ApJS...61..531S,1987ApJ...317..781B,1987ApJ...314..272Y}.
After subtracting
background, the resulting excess equivalent widths were converted to H$\alpha$ luminosities. It was found that rotation does not play a role in
the degree of H$\alpha$ activity, i.e., the stars are not in the spin-down
phase so they must be quite young. The H$\alpha$ luminosities appear to be
saturated at a fixed fraction, 20\%, of the coronal X-ray luminosity
which may be their ultimate energy source or indicative of the same
heating mechanism in both spectral domains. Our results were the first
to describe the nature of the saturated chromospheric state.

\section{The Retirement Years}

\subsection{Early period (circa 2001)}

I became interested in the effects of rotational distortion on the
internal structure of solar-like stars during conversations with Steve
Jackson (PhD, U.\ of Chicago) who was at HAO at the time.
I learned that he created the first code to successfully solve the
coupled stellar interior equations and the Poisson equation for the
gravitational potential because of the non-spherical shape of the
density profile. The numerical methodology was based on a Princeton
(Ostriker's) scheme (labeled SCF) of iterating between individual
sequential solutions for the interior and potential equations. The code
worked very well for rapidly rotating early spectral type (massive
stars).

\begin{figure}
    \centering
\includegraphics[width=.98\textwidth]{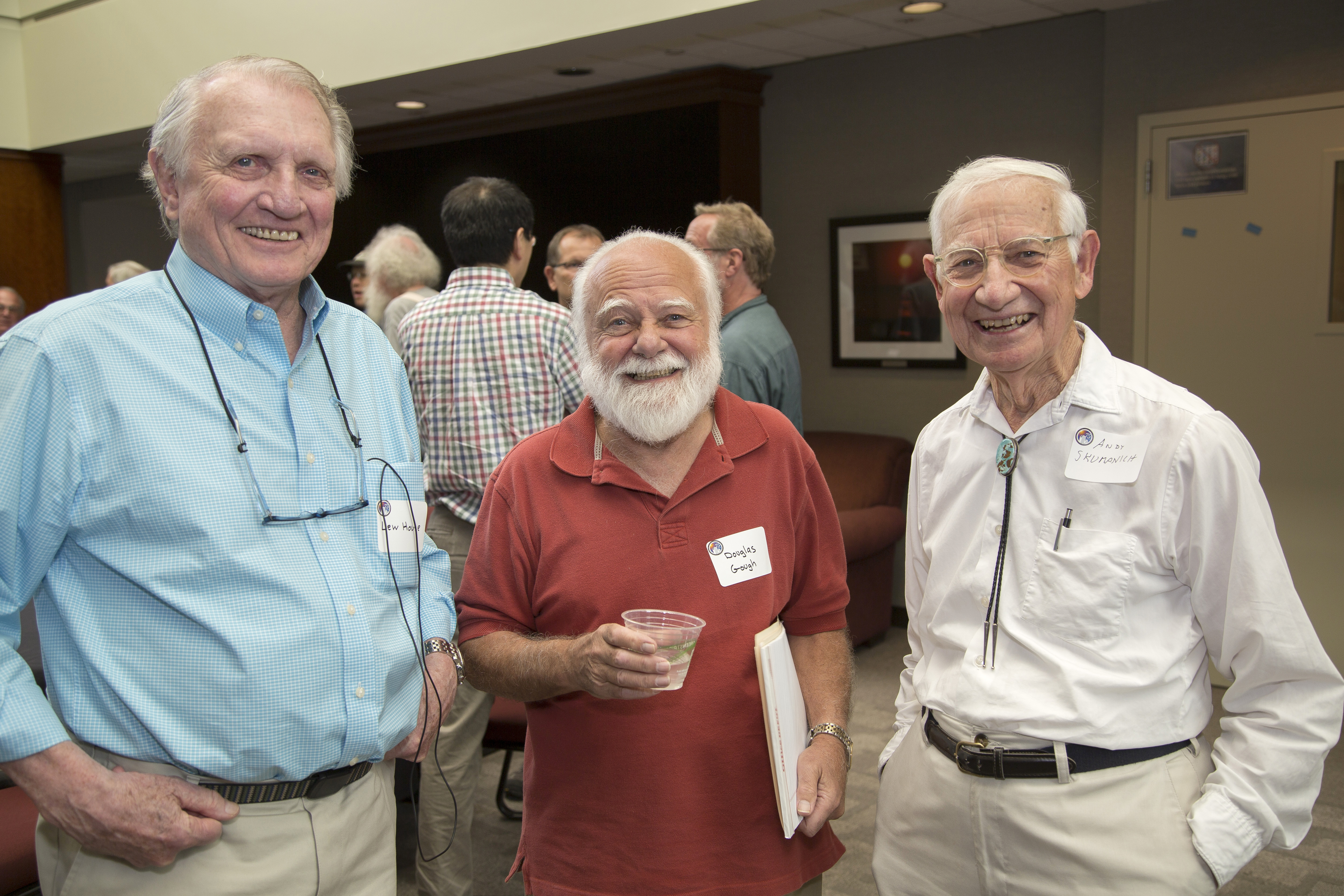}
    \caption{Attendees to 75th Anniversary Celebration of HAO (2015). Left to right, L. House (HAO), D. Gough (U.\ Cambridge, UK) and A. Skumanich (HAO).}
    \label{fig:HAO75}
\end{figure}

Unfortunately the code could not be used to study solar-like stars since
it failed to converge for stars with masses less than 9 M$_\odot$.
The methodology had to be reformulated.

Steve's complained to me that his attempt to revise the SCF procedure
was stymied by a lack of computer resources to test his progress.
When I discussed the desirability of having a revised version of Steve's
code at HAO with Keith McGregor (Head of the Stellar Interiors Section)
he agreed and arranged for a visitors appointment for Steve.

To our surprise, Steve arrived in Boulder with a punch-card version of
his code and associated subroutines (a vestige of his work being done at
Princeton -- I recall my own experience there). A card-reader was found
at one of the older federal labs in Boulder, and Steve transferred his
extant code to a SUN workstation. He began testing various
reformulations and finally arrived at a new successful SCF version that
converged for any stellar mass.

We explored the nature of these solutions across the two-dimensional
parameter space (rotational amplitude and gradient) and compared the new
SCF solutions with those from other extant methods. The result was a
robust computational algorithm 
\citep{2005ApJS..156..245J}, with updated auxiliary
physical processes, such as the equation of state, opacity subroutines,
energy generation reactions, convective energy transport, chemical
composition, and so forth. The new code was able to treat distortions
not attainable by other methods.

The first precise interferometric measurement of the shape of $\alpha$ Eri
(Achernar, a B star) provided our first opportunity to model the physical
structure of a rotationally flattened star. Employing masses for
main-sequence stars of mid- to early-B spectral type, we could
reproduce Achernar's inferred equatorial and polar dimensions. This was
achieved through a combination of rotational flattening/distension and a
suitable inclination of the rotation axis \citep{2004ApJ...606.1196J}.

Despite these successes, the models were discrepant in other respects:
being (on average) cooler and more rapidly rotating than indicated by
the observations.

Finally, we calculated a number of models for chemically homogeneous,
differentially rotating, main-sequence stars with masses in the range
1-2 $M_\odot$ \citep{2007ApJ...663..560M}. For a rapidly rotating Sun, we found a
reduced radiative luminosity. Relative to non-rotating stars of the same
mass, all of our rapidly-rotating models exhibited reduced luminosities
and effective temperatures, and they displayed a flattened photospheric
shape (i.e. decreased polar radii).

Thus solar-like stars not only brighten because of chemical evolution
(nuclear transmutation), but also because of rotational spin-down. For a
fixed ratio of axial to surface equatorial rotation rates, increased
rotation typically deepens convective envelopes, and shrinks convective
cores. It may also lead to a convective core ($M = 1\,M_\odot$) or envelope
($M = 2\,M_\odot$). These features are absent in a non-rotating star of the
same mass.

\begin{figure}
    \centering
\includegraphics[width=.8\textwidth]{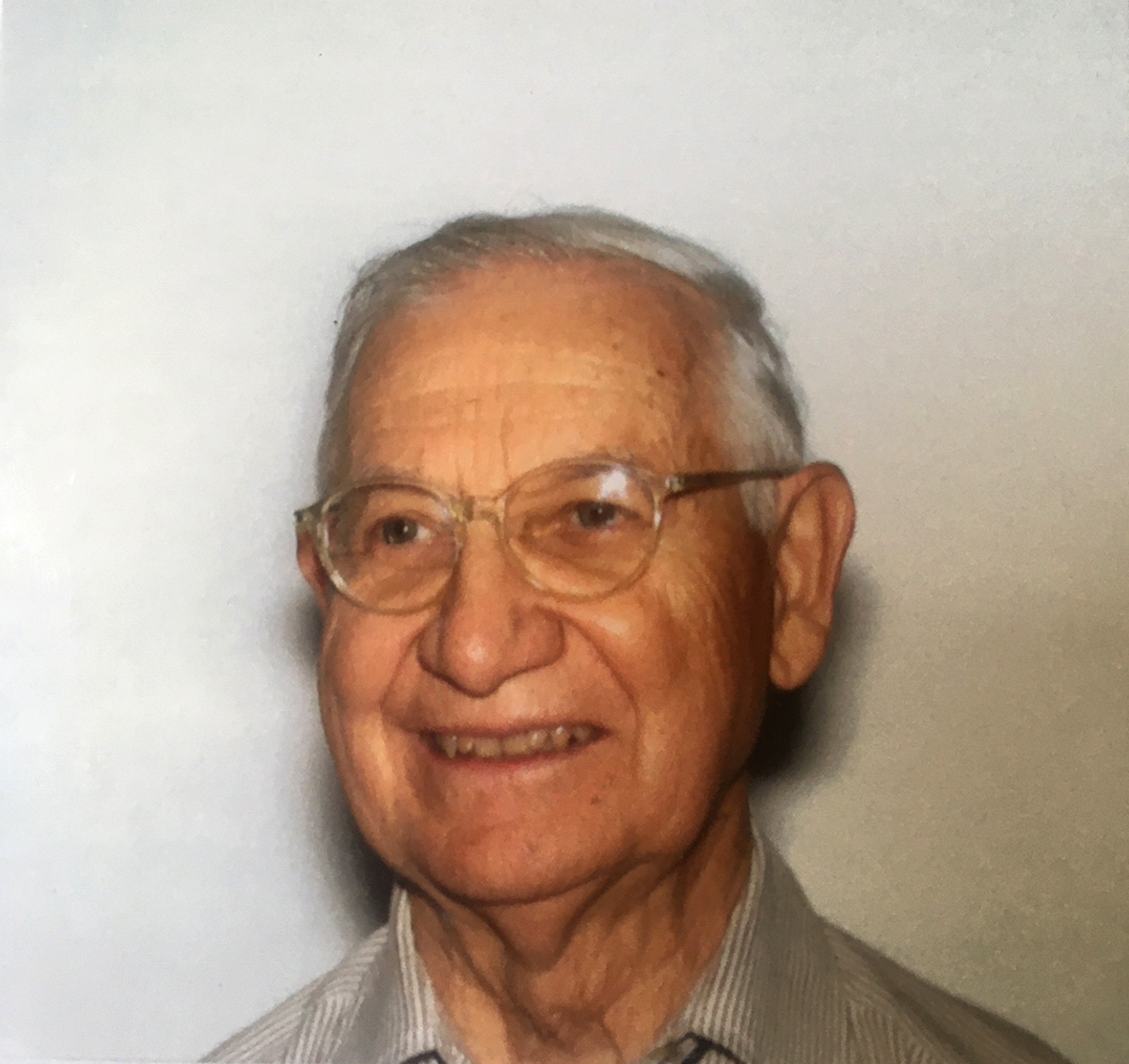}
    \caption{A. Skumanich, 2009 (80 years)}
    \label{fig:Sku_80yrs}
\end{figure}

Theoretical studies by D.~R.\ Reese (Department of Applied Mathematics,
University of Sheffield, UK) suggested that one might use
asteroseismology, the study of stellar pulsations, to probe the internal
structure of rapidly rotating stars. This permits the observed pulsation
spectra to constrain/test theoretical models. We provided several SCF
models for Reese to study with his pulsation code  \citep{2009A&A...506..189R}. A successful identification of pulsation modes for a rapidly-rotating $M = 25\, M_\odot$ model was found \citep{2009A&A...506..183R}. This advance has opened new avenues for asteroseismology.

Another highlight of my later years at HAO was the celebration of the 75th Anniversary 
of the observatory, where I had the good fortune of meeting again with old friends and colleagues (Figure~\ref{fig:HAO75}).

\subsection{Late Period (2017) -- Solar Physics Handbook}

In December of 2016, my wife, Mary, and I had the occasion to enjoy a
reunion with our French friend and colleague Jean-Claude Vial (see Sect.~\ref{sec:OSO-8}) in San Francisco, during our customary winter stay in
California. He mentioned a handbook that he and another colleague,
Oddbjørn Engvold, were proposing to the publisher, Elsevier. Because of
my work spanning both solar and stellar astrophysics, he invited me to
join them. The rationale for the project is quoted here from a January
2017 email to me from Jean-Claude: \emph{``The aim of this book is to present an up-to-date view of the
entire field of solar physics for illustration of the significance of
the Sun as a guide star in stellar astrophysics. The nature and physical
processes, which were first revealed and studied in the Sun, shed
clarifying lights on observed characteristics of various type stars.
Some methods first used in solar physics were later applied in stellar
physics. [...]  In the event that you would find this book project meaningful and
worthwhile, would you consider joining us as co-editor?''}

I responded on February 12 that the idea of such a book as he
described seemed interesting and would be a resource to students and
professionals alike and that I had decided to be a part of the effort.
We prepared a proposal and sent it to the editor at Elsevier. It was
refereed, and after some changes, accepted and ultimately published \citep{2019sgsp.bookD..17E}. The chapters and their
authors are listed here:\vspace{6pt}

\noindent
\textbf{The Sun as a Guide to Stellar Physics}\\
Edited by Oddbjørn Engvold, Jean-Claude Vial, and Andrew Skumanich
\begin{enumerate}[leftmargin=24pt,itemsep=-3pt]
\def\labelenumi{\arabic{enumi}.}
\item
  Discoveries and Concepts: The Sun's Role in Astrophysics\\
  \emph{Jack B.\ Zirker and Oddbjørn Engvold}
\item
  Stellar and Solar Chromospheres and Attendant Phenomena\\
  \emph{Tom R.\ Ayres}
\item
  The Sun's Atmosphere\\
  \emph{Alexander I.\ Shapiro, Hardi Peter, and Sami K.\ Solanki}
\item
  Helioseismic Inferences on the Internal Structure and Dynamics of the
  Sun\\
  \emph{Sarbani Basu and William J.\ Chaplin}
\end{enumerate}
\noindent
\textbf{\emph{Atmospheric Structure, Non-Equilibrium Thermodynamics\\ and Magnetism}}

\begin{enumerate}[leftmargin=24pt,itemsep=-3pt]
\def\labelenumi{5.\arabic{enumi}.}
\item
  Spectroscopy and Atomic Physics\\
  \emph{Philip G.\ Judge}
\item
  Models of Solar and Stellar Atmospheres\\
  \emph{Petr Heinzel}
\item
  Spectropolarimetry and Magnetic Structures\\
  \emph{Kiyoshi Ichimoto}
\end{enumerate}

\begin{enumerate}[leftmargin=24pt,itemsep=-3pt]
\def\labelenumi{\arabic{enumi}.}
\setcounter{enumi}{5}
\item
  Coronal Magnetism as a Universal Phenomenon\\
  \emph{B.~C.\ Low}
\item
  Magnetohydrodynamics and Solar Dynamo Action\\
  \emph{E.~R.\ Priest}
\item
  Solar and Stellar Variability\\
  \emph{Marianne Faurobert}
\item
  High-energy Solar Physics\\
  \emph{H.~S.\ Hudson and A.~L.\ MacKinnon}
\item
  Space Weather at Earth and in Our Solar System\\
  \emph{No\'e Lugaz}
\item
  The Solar- Stellar Connection\\
  \emph{Gibor Basri}
\end{enumerate}
\textbf{\emph{Instrumentation}}

\begin{enumerate}[leftmargin=24pt,itemsep=-3pt]
\def\labelenumi{12.\arabic{enumi}.} 
  \item
    Observations of the Sun From Space\\
    \emph{Alan Title}
  \item
    High-Resolution Ground-based Observations of the Sun\\
    \emph{Oddbjørn Engvold and Jack B.\ Zirker}
\end{enumerate}

\begin{enumerate}[leftmargin=24pt,itemsep=-3pt]
\def\labelenumi{\arabic{enumi}.} 
\setcounter{enumi}{12}
\item
  Solar Data and Simulations\\
  \emph{Neil Hurlburt}
\item
  Challenges and Prospects for the Future\\
  \emph{Jean-Claude Vial and Andrew Skumanich}
\end{enumerate}

\subsection{The Last Hurrah (2019)}

The last paper of my career \citep{2019ApJ...878...35S} was published 10 years after my prime age (Figure 7). It arose from editing Tom
Ayres's contribution to the Handbook (see Section 11.2). Ayres, in discussing rotational
spin-down, displayed an explicit interpolation formula for the torque
derived from numerical solutions of a magnetized solar wind for
different values of the wind parameters, magnetic field and mass-loss
rate \citep{2012ApJ...754L..26M}. 

The torque formula contained the product of two known monomial (power
law) functions of each of the wind parameters. To obtain a solution for
the temporal variation of the angular momentum one must know how they
vary with the rotation rate.

I contacted Sean Matt for his advice regarding coronal wind theory and
rotational spin-down. I also mentioned that, in the absence of a theory
predicting the dependence of these parameters on rotation rate, one
could derive their dependence by imposing two observationally derived
rules governing the spin-down. He thought my idea would make for an
interesting paper, so he encouraged me to develop the argument. He
provided insightful comments and valuable suggestions at various stages
of my analysis.

The torque parameters are assumed to depend on the rotation rate by
monomials with different exponential indices. In the case of the
magnetic field, I had recourse to the relation implied in my popular
1972 paper \citep{1972ApJ...171..565S}. I found that the chromospheric Calcium
luminosity (which depends on the global magnetic field) was proportional
to the rotation rate. By representing the chromospheric emission by an
unknown power law of the field strength, i.e., with an unknown
exponential index, one obtains an explicit form for the field factor in
the torque.

For the mass-loss factor I assumed that it varied with a monomial
function of the field with an unknown index. This led the torque to have
a cumulative index for its monomial dependence on rotation. By imposing
the Skumanich Law requirement one obtains a linear relationship between
the two indices. If one has an observational value for either index one
knows the value of the other.

The index-pair was calculated for twenty-five known activity
correlations. The wide distribution of points along the solution line in
the plot of one index against the other indicated a lack of consistency
among the activity correlations. This is probably due to observational
data errors or data reduction errors (e.g., incorrect baseline or
zero-point corrections) or non-global sources.

The box-average of the distribution implies that the global surface
magnetic field of a solar-like star scales linearly with rotation while
the mass-loss rate scales with the square of the global field strength,
i.e. with the magnetic energy density in the system.
The first conclusion reflects on the dynamo field and its emergence
while the latter has implications for the energetics involved in driving
the wind.

\section{Academic Engagement through my Career}

\subsection{Faculty Courses}
When I joined HAO, I was also appointed as an adjoint lecturer and
finally professor in the CU Boulder graduate program of
Astrogeophysics. Later I was appointed as professor in the Physics
department.

At the behest of Mahinder Uberoi, Chair of the Aerospace Engineering
Department at CU Boulder, who wanted his students to have some background in
Astrophysics, I devised a first year Modern Astrophysics course, Physics
580, open to students in Astrogeophysics, Physics and Aerospace
Engineering.

I was invited to lecture at the August 1976 Erice summer school (Sicily)
where I discussed the role of escape probabilities in estimating `back
of the envelope' source functions, often to check the convergence of
more elaborate numerical calculations. The nature of radiation operators
and their matrix representation was also presented.

Among the students auditing my lectures was Ester Antonucci who was very perceptive and interactive with regard to the materials I presented. She was at the school to see if Solar Physics was a possible career path. Perhaps my lectures helped her decide that it was. We soon became friends and colleagues. I was pleased to recently learn that she was chosen by a Solar Physics panel as a memoirist and published her Memoirs last year \citep{2022SoPh..297...89A}.

\subsection{Doctoral Candidates}

I have benefited both scientifically as well as personally from working
with graduate students as either their PhD advisor or as a mentor. I was
involved with the following students.
\vspace{6pt} \par \noindent
1) Loren Wallace Avery entered (1963) the Department of Astrogeophysics at CU Boulder as a
PhD student, and selected Lewis House as his thesis advisor. House
suggested that Avery consider the use of the Monte Carlo method to
determine the internal radiative field for a resonance line (two level
atom) in open cylindrical geometries for the self-excited case with and
without external illumination (see Sect.~\ref{sec:RT} for a discussion
of the formation of resonance lines).

With my knowledge of Monte Carlo methods I was invited by House to
participate in Avery's PhD candidacy as a mentor and co-chair on his
thesis committee.

With our tutelage Avery developed a Monte Carlo program to determine the
source function for a level-atom for cylindrical geometries. The
resulting thesis, titled ``A Monte Carlo Calculation of Radiative
Transfer in Cylinders with Application to Solar Spicules,`` was accepted
and his degree was awarded in 1969.

Avery's work on the scaling of the source function for a variety of
sizes appears in \cite{1969JQSRT...9..519A}. His results
with the Monte Carlo method confirmed my results derived by the standard deterministic method \citep{1967AJ.....72T.828S}.
\vspace{6pt} \par \noindent
2) Gary J. Saliba was a Doctoral candidate at the Department of Applied
Mathematics (University of Sydney, Australia) when I met him during my
first appointment to the department in 1983.

His thesis title was ``Non-LTE Scattering Resonance Polarization in
Solar Spectral Lines''. The thesis was accepted and his degree was
awarded in 1986. I was involved only as a mentor sharing my
understanding of the resonance scattering process and discussing various
aspects his analysis.
\vspace{6pt} \par \noindent
3) Harrison P. Jones entered as a PhD student in the Department of
Astrogeophysics and started to work with me on generalizing the \cite{1967AnAp...30..669A} flux-divergence formalism for the
statistical equilibrium of a two-level atomic model to a non-iterative
scheme applicable to multi-dimensional media based on the method of
characteristics \citep{1968rla..conf...79J}.
The conversion of
the resultant flux-divergence operators to their algebraic forms appears
in \citealt{1973ApJ...185..167J}.\footnote{Cited as recently as 2022.}

Jones incorporated these results in his thesis ``Line Formation in
Multi-Dimen{-}sional Media'' which was accepted and his degree was awarded
in 1970. The thesis was published as NCAR Cooperative Thesis No. 21 by
University of Colorado and the High Altitude Observatory, NCAR.
\vspace{6pt} \par \noindent
4) Philippe Lemaire (Associate Scientist, LPSP) came to HAO in the fall
of 1970 to study the formation of resonance lines with me. He had
designed and executed a balloon experiment to measure the solar Mg II h
and k resonance lines at high spectral resolution. We derived one of the
first models of the \ion{Mg}{2} chromosphere \citep{1973A&A....22...61L}. This work was ultimately included in his PhD thesis ``Recherches
sur l'émission de la chromosph\`ere solaire dans les raies du magnésium
ionisé'' which was accepted in June 1971. I was an invited member on
his thesis committee in Paris.
\vspace{6pt} \par \noindent
5) Michel Hersé, a graduate student at the Université Pierre et Marie Curie (Paris VI), was a scientist assistant to Jacques E. Blamont at Service d'Aéronomie du Centre National de la Recherche Scientifique (CNRS). I had met Michel in 1973 during my sabbatical stay at LPSP. He was working on high resolution pictures of solar bright ‘grains’ in the ultraviolet  as observed from a ballon. At this time I was working on such bright ‘grains’ as observed in the Ca II K emission line \citep{1975ApJ...200..747S}. I was soon invited by Blamont, Président of Michel’s thesis panel, to participate as an advisor. Michel received his Docteur ès Sciences degree in 1976 with the thesis “Structure Fine du Soleil dans l’Ultra-Violet”. We have remained friends and colleagues since.
\vspace{6pt} \par \noindent
6) Benedict (Ben) Domenico became a graduate student in the Department
of Physics and Astrophysics and started to work with me to convert the
statistical population equation with flux-divergence operators for a two-level atomic model \citep{1967AnAp...30..669A} to a multi-level model that allows
for subordinate lines (non-ground state transitions). The now vector
population equations were solved by applying an inversion procedure
based on quasi-linearization 
\cite[Newton-Raphson;][]{1968rla..conf..475S}.

This system was successfully tested and used to analyze the formation of
Hydrogen Lyman $\alpha$ and Lyman $\beta$ resonance lines as well as Balmer H$\alpha$ for the solar atmosphere \citep{1971JQSRT..11..547S}.

Domenico adapted the new vector population algorithm in his thesis, ``On
the Application the Generalized Newton-Raphson Method to the
Singly-Ionized Calcium Line Formation Problem In Model Stellar
Atmospheres'' where he investigated the formation of the two resonance
\ion{Ca}{2} H and K lines as well as the \ion{Ca}{2} infrared triplet (IRT), and the
ground state of \ion{Ca}{3}. In addition he studied the implications of Olin
Wilson's observations \ion{Ca}{2} emission stars on the derived thermal
models. The thesis was accepted and his degree was awarded in 1972. The
thesis was published as NCAR Cooperative Thesis No.~25 by University of
Colorado and the High Altitude Observatory, NCAR.
\vspace{6pt} \par \noindent
7) Graham A. Murphy, another doctoral candidate at the Department of
Applied Mathematics (University of Sydney, Australia), who became a
research assistant to the `Gang of Three' (Rees, Lites, Skumanich). One
of his tasks was to develop a method to integrate the Stokes vector
equation of transfer that would yield the emergent Stokes vector for a
given model atmosphere 
\citep{1987ApJ...318..930L,1988ApJ...330..493L}. This became a part of his
thesis ``The Synthesis and Inversion of Stokes Spectral Profiles''. It
was published as Cooperative Thesis No.~124 by the University of Sydney
and the High Altitude Observatory, NCAR, in 1990.
\vspace{6pt} \par \noindent
8) Arturo L\'opez Ariste, a graduate student at l'Universite Paris 7, Denis Diderot, was a research assistant to Meir Semel, his thesis
advisor, at DASOP, Observatoire de Paris, Section de Meudon. I had met
Arturo in 1995 when I joined Semel at DASOP. I was soon invited to join
Ariste's thesis committee as an advisor. He received his PhD degree in
October 1999 with a thesis entitled ``La Spectropolarimetrie en
Astrophysique: Application au diagnostic des champs magnétiques solaires
et stellaires''. After his degree he came directly to HAO in January
2000 as a post-doc and then in 2001 was appointed Scientist~I on the
regular staff. Eventually, he took an appointment at CNRS, France.  We continued our relationship as friends and colleagues.

\section{Additional Collaborations through my Career}

\subsection{Durney, Bernard (HAO)}

The \cite{1968ApJ...152..255D} paper was the first to study
non-radial adiabatic oscillations of slowly rotating polytropes. Using a
spherical harmonic expansion method they identified structures that are
now called gravitational gyroscopic waves or gyres. This work has been
referred to as ``pioneering'' in recent publications \cite[see, e.g.,][]{2013MNRAS.435.3406T}.

\subsection{Poland, Arthur (HAO)}

In order to provide thermodynamic models for use in the analysis of the
hydrogen Balmer H$\alpha$ observations in absorption or emission (and
alternatively, lines from minority species), from spicules and
prominences, \cite{1971SoPh...18..391P} calculated the
ionization and excitation equilibrium for hydrogen in a variety of slab
model atmospheres that are irradiated from both sides by photospheric,
chromospheric, and coronal radiation fields. Such vertically standing
structures were used as a 2D representation of their form. This was
possible because the major photon escape route, which fixed the internal
radiation field, is fixed by the slab thickness. The flux-derivative
version of the statistical equilibrium equations proved to be efficient.
The high citation number for this publication indicates the models were
very useful.

\subsection{Liu, Sou-Yang (HAO)}

Sou-Yang arrived at HAO in 1973 for a post-Doc appointment with a
high-resolution observational data set dealing with the fast time
evolution of the \ion{Ca}{2} K line. The observations refer to internetwork
1-arcsecond bright grains in a quiet region at the center of the disk

The data indicated a wavelength localized brightness enhancement in the
wings of the line that moved closer to line center (hence upwards) and
ended with a blue shifted emission in the Doppler core of the line \citep{1974ApJ...189..359L}.

Liu wanted to learn how to apply the HAO radiative resonance line
inversion algorithms so I instructed him and joined him to determine the
thermal and  dynamic model atmospheres needed to fit the spectral line
data for several specific temporal line profiles
\citep{1974SoPh...38..109L}. I proposed that we consider a rest thermal state for the
upper photosphere and lower chromosphere  that is then affected by a
thermal and dynamical perturbation whose height and width changes with
time. The resulting inversion yielded a downward flow containing an
upward moving wave. These results implied that the grains were an
acoustic phenomenon.

This was demonstrated ultimately with a hydrodynamical simulation of the
grains by \cite{1997ApJ...481..500C}, who
showed that upward propagating acoustic waves convert non-linearly into
shocks in a downflowing stream and produced cell grains.

\subsection{Jones, Harrison P. (NASA-GSFC)}

Harry, who was now  at the NASA Goddard - Southwest Solar Station, and I continued our association  and considered the general the effects of a finite geometrical structure on
the internal radiative field density,  \cite{1980ApJS...42..221J}. We showed that the volume-averaged
escape probabilities, which are closely related to radiation loss rates,
\emph{scale} with the mean chord length (thickness $D$ for a slab, radius
$R$ for a cylinder), and thus the radiative losses are approximately the
same in the two geometries. A limitation of this result is that they
are only strictly valid when the absorption coefficient per unit length
is constant with geometric position

This geometric scaling law approach was confirmed by the 2D NLTE
computations of \cite{1978ApJ...220.1001M} with a two-level atom,
and later applied to prominences for the case of the \ion{H}{1}, \ion{Ca}{2} and \ion{Mg}{2}
lines by \cite{1982ApJ...253..330V}.

\subsection{Lean, Judith (CIRES)} 


While at CIRES,\footnote{Cooperative Institute for Research in Environmental Sciences (CIRES), University of Colorado/NOAA, Boulder, E.~O.\ Hulburt Center for Space Research, Naval Research Laboratory, Washington, D.C. } Judith, with whom I had collaborated on the development of the solar \ion{Ca}{2} K
irradiance model (see Section 10.1), continued our collaboration which
included a variety of others from the atmospheric irradiance `cartel'.
In addition to estimating the UV  and Lyman $\alpha$ irradiances   for
terrestrial atmosphere analysis \citep{1982JGR....8710307L,1983JGR....88.5751L}, 
our studies included the nature of the Maunder
Minimum \citep{1992GeoRL..19.1591L,1995GBioC...9..171L}, as well as the issue of a
non-cycling state \citep{1992PASP..104.1139W}. With the exception of \cite{1995GBioC...9..171L}, all these works are
fairly well cited into 2022.

\subsection{Semel, Meir (DASOP)}

In the Fall of 1995, I took a leave of absence to join Meir Semel, whom
I met during my early sojourns to France, at the D\'epartement
d'Astronomie Solaire de l'Observatoire de Paris (DASOP) in Meudon to study
the nature of the radial current densities (in the heliocentric
coordinate frame, HCF) associated with magnetic field regions on the Sun
as observed with the ASP polarimeter. These currents yield important
boundary conditions for the outer solar atmosphere.

We note that the following methodology \citep{1998A&A...331..383S}
requires one to assign the observer's frame magnetic field to both a
parity free line of sight (LOS) component and a parity dependent
(perhaps arbitrarily assigned) plane of the sky (POS) component. These
are then mapped to the HCF and then projected on to the solar tangent
plan. Both fields yield a radial current via Amp\`ere's law. These
currents are then treated separately and differently.

In the case of the radial POS-current an algebraic method was used to
modify the current into a sum of two quadratic forms with parity
dependent coefficients but whose absolute value was independent of any
particular azimuth disambiguation (parity assignment) procedure. This
yielded an invariant or `free' current which was used to replace the
parity dependent form in other expressions.

The use of this construction yielded a smoother radial current than that
derived from applying Amp\`ere's law to the POS field derived by using the
disambiguation from HAO-AZAM.\footnote{A user-interactive azimuth
  assignment procedure.} It was found that similar current
discontinuities occur in both maps, confirming the correctness of the
new construction, but whose location was found to be occasionally
disambiguation dependent.

The final construction adds the parity free radial current derived from
the LOS field to our invariant current in such a way so as to yield two
independent and disambiguated currents, one of which represents the
smallest possible current allowed by the data. The selection of the
minimum current disambiguates any particular magnetic map at any
position on the solar disk.

We applied this minimum current disambiguation \citep{1996SoPh..164..291S} to two HAO magnetic maps with different positions on the disk of
the active region NOAA 7201 (see Sect.~\ref{sec:MFE} above). The site of interest contained an emerging delta spot.

A comparison of our minimum current density map with that derived with
Amp\`ere's law applied to the AZAM maps showed that either disambiguation
method yielded quite similar results. Our construction yielded strong
linearly extended features that appear on all maps near the neutral line
in the plage and delta spot region. They are not of uniform amplitude
along their length, i.e. they are often fragmented.

Note that long after its introduction, the value of our
parity-free absolute-radial-current density method was well demonstrated by its
use by Georgoulis\footnote{As presented in \cite{2006SoPh..237..267M}.} as part of the initial current in an iterative
disambiguation scheme. It significantly improved the ultimate accuracy
of his disambiguation of synthetic data.

Considering the complex transformations involved in the paper, the
unusual publication history of \citep{1998A&A...331..383S} was not surprising. It was
submitted to A\&A on July 11, 1996, accepted July 24, 1997 and published
in 1998.

\subsection{Leka, K.~D.\ (NCAR-ASP)}

Leka came to HAO in 1995 via her appointment as a Fellow in the Advanced
Studies Program at NCAR. She expressed an interest in the ASP science
program so I became her scientific guide and tutor in the various
subjects associated with Stokes polarimetry. Her first Stokes data
reduction adventure was with the Stokes observations of the so-called
quiet (no ``activity'') region at the center of the solar disk. This led
to the discovery of `HIFs', transient small-scale horizontal
internetwork fields as presented in \cite{1996ApJ...460.1019L} (see 
discussion in Section ~\ref{sec:unobserved}).

I suggested to KD that we consider the temporal evolutions of fine
structures and magnetic fields around pores. The consequence was \cite{1998ApJ...507..454L}, one of the first quantitative studies that set
the basis for the development of the field; refer to discussion in
Section 9.3.

Our final collaborative work was to derive a suitable method to
study the nature of an active region’s helicity coefficient, $\alpha$AR,
which is defined by the ratio of the photospheric radial current
(density) by the photospheric radial magnetic field. The result
was the paper by \cite{1999SoPh..188....3L}.
We tested three methods to calculate $\alpha$AR and discussed their
limitations and examined the influence of data noise on their
results. The discrepancies, agreements, and overall robustness
of the different methods are also discussed. The paper was well
received in the coronal activity field and  has been frequently referenced up to this day.

\subsection{L\'opez Ariste, Arturo (HAO)}

In thinking about the new Stokes inversion method based on applying
Principal Component (PC) analysis to a synthetic and observed Stokes
data base I realized that one could obtain an unbiased estimate of the
physical content of each Stokes parameter profile by applying the
Principal Component or Orthogonal Decomposition directly to the spectral
covariance matrix (using wavelength pairs) for the Stokes intensity and
net polarization spectral profiles, at any position. This matrix is then
diagonalized by singular value decomposition to obtain the
eigenfunctions (orthogonal components) and eigenvalues.

The PCA method expands the Stokes intensity and net polarization
spectral profiles, at any position, in terms of these eigenfunctions.
The expansion is ordered by the magnitude of the relevant eigenvalue
from largest to smallest. The ordering represents a perturbation
expansion

Arturo volunteered to use his PC codes to help me with my project, which
appeared in \cite{2002ApJ...570..379S}, one of the last
papers of my career.

We found that the ordering allowed us to examine the physical content of
the first few orders of the set of 40,000 profiles for each Stokes
parameter for a solar active region \cite[see][OSO-8 spot]{1987ApJ...322..473S}.

In particular, the analysis showed that the eigenvalues found by the PCA
construction can be related to certain physical quantities, e.g., to the
line-of-sight velocity and the magnetic Zeeman splitting for the
unpolarized Stokes $I$ profiles, to the transverse field components for
the linearly polarized Stokes $Q$, $U$, and the longitudinal magnetic flux
for the $V$ profiles.

Note that this quick-look method easily found that the bright
quiet-Sun points (grains) have an upflow signature, while the dark
regions have a downflow-one, in good agreement with that derived by \cite{1974SoPh...38..109L}.

The most recent reference to this Quick-look PCA-spectral method was a
publication by \cite{2022MNRAS.514.2333L}. They adopted the method for
the determination of stellar magnetic fields from a time series of
observations of the stellar circular or Stokes~$V$ polarization profiles.

They concluded that the method allows an easy first glance at key
parameters of the stellar magnetic field topology before more advanced
techniques like Zeeman-Doppler-Imaging (ZDI) are used to reconstruct a
full map of the surface magnetic field. It can be conveniently applied
to a large number of stellar targets.

\section{The 50th Year Anniversary} \label{sec:50th}

The influence of my 1972 paper \cite[][hereafter Sku72]{1972ApJ...171..565S},
which studied the age dependence of surface rotation state (magnetic
wind braking) and chromospheric luminosity (\ion{Ca}{2} emission,
signature of magnetic activity) has persisted since its publication.

It has spawned many studies relating rotation periods to magnetic
activity for other mass stars, e.g., by \cite{1984ApJ...287..769N}, and many
other studies. A new field of Gyrochronology was inspired which uses
surface rotation rates for the first time as an age diagnostic \citep{1999A&A...348..897L}. The advent of the IUE satellite
data (c.~1980) allowed one to search for Sku72-like relations in the UV and
X-ray domain. Likewise with the Kepler satellite (c.~2016), the studies
of young and intermediate seismic ages confirm the Sku72 results for solar
analogs in general except for stars older than 2~Gy.

The current influence is well illustrated by the range of papers
presented at a 2022 meeting to celebrate the 50th anniversary of the Sku72
publication date (\url{https://skumanich.wdrc.org}). The celebration was initiated by Travis Metcalfe and
colleagues, and hosted as an in-person
conference organized around the major science themes of the Sku72 paper.
The announcement notes that \emph{``the paper is one of the most impactful
publications in the history of HAO, with more than 1500 citations to
date''.}

The Festschrift abstracts for all of the talks and posters presented at
the meeting, and the presentations
themselves have been archived
for posterity.\footnote{\url{https://drive.google.com/drive/u/0/folders/153u2Ap8ypL3caAJRvaXItTZL1NNeOpXr}.} My own presentation, ``The discovery path of the inverse square root of age relations for solar-type stars'', with explanatory slides, can also be downloaded there.\footnote{\url{https://docs.google.com/presentation/d/1hd7jFERZqrHSkUNk5S-Y6YNXDxfRrOGm}.}

Holly Gilbert, the Director of HAO, opened the meeting with a welcome and
wishes for a fruitful celebration. After a few words commenting on my long
association with the observatory she invited me to the front to be
presented with a framed poster by the American Astronomical Society
congratulating me \emph{``On the fiftieth anniversary of the publication
[...] in the Astrophysical Journal. This famous and highly cited paper has had a profound influence
on the development of stellar astrophysics in the intervening years.''}. I
was surprised and quite pleased with the AAS recognition.

At the celebration, I was excited to see and met some of the young
people contributing to the advancement of Solar-Stellar Astrophysics.
After 20 or so years of retirement I was also overcome with emotion on
meeting many dear former colleagues and friends again. ``Those were the
days my friends.''

\section{Personal Reflections}

My scientific career path has been an eclectic one. With Martin
Schwarzschild's tutelage I developed an interest in mathematical
analysis and model building, with Alexander Nikolaevich Vyssotsky, an
interest in the relationships or correlations between observations. This
was, to some extent, the fulfillment of my childhood curiosity and need
to understand. Published work of, and interaction with, peers and
collaborators often opened other possible paths to explore. You will
find such examples in this memoire. I had no specific research plan when
I was ready to enter the work force. My direction was driven by the
available opportunities.

One feature of my early scientific career was my involvement at LASL to
do physics experiments with high energy density sources as provided by
nuclear weapons. To have witnessed the detonations of atomic (fission)
and thermonuclear (fusion) sources was an awesome experience. The
fission explosions turned night (predawn) into blue daytime that lasted
a fair number of heart beats while the fusion explosion `sunrises'
seemed to last fearfully longer. Once I left LASL I would have an
infrequent dream that while walking along a street there would be a
flash of light above me and, recognizing the BOMB, I would drop to the
curb, cover my head and after the shock passed would rise and start
brushing dust off my clothes \ldots\ as I imagined the Japanese fishermen
did during the first H-bomb event.

On the occasion of a stellar \ion{Ca}{2} observing session at Mt Wilson with
Olin Wilson I had a night time opportunity to climb (furtively) to the
top of the 100-inch dome and behold the luminous carpet of Los Angeles
below me. In my fancy the curved dome became Earth, and I was Atlas (in
reverse). It was an unforgettable image. Heights were never a problem
for me. At Las Alamos I worked on a 500-ft tall open tower in the case of the
neutron PINEX experiment.

My interest in Art and History, which developed in my youthful years, was augmented by visits to museums and historical sights in different US cities where I attended professional conferences. Conferences abroad proved to be even more enriching,  given the birth of the Renaissance in Europe. Thus my 1-year sabbatical at LPSP (1973-1974) in the environs of Paris and my semester long stays at the University of Sydney (1984, 1986) and University of Florence  (1988, 1993) were fulfilling.  During our year-long stay in Versailles, my family and I were able to do the fabled ‘Grand Tour’ that was the apex of a humanistic education.

I have had extraordinary luck and unexpected changes -- good breaks --
in my life line at various stages of my life. My greatest satisfaction
was to be able to work with outstanding colleagues, post-docs, grad
students, and scientific staff assistants. My education was possible
through the unexpected assistance from my `village', personal labor,
librarians, and teachers who cared, college and university endowments
and ultimately, through government expense. I am grateful for all of the
above and very thankful to have lived in a society and at a time when
the pursuit of knowledge was considered a worthy direction of community
resources.

\begin{acks}
The author is first and foremost grateful to Roberto Casini (HAO-NCAR)
for converting the initial manuscript into \LaTeX, implementing the bibliography, and for overall editorial assistance. He is also grateful to Michael Kn\"olker and Bruce Lites (HAO-NCAR, emeriti), Tom Bogdan (NSO), Arturo López Ariste (U.\ Toulouse, France), and an anonymous referee for helpful comments and corrections to this memoir. He thanks Holly Gilbert
(HAO-NCAR, director) for her support, and Wendy Hawkins (HAO-NCAR) for her help with the
included pictures. NCAR librarians Michael Flanagan, Krista Gawlowski, and Laura Hoff are
graciously recognized for their help in providing many of the bibliographic sources.
\end{acks}


%

%

%
%

%
%
\bibliographystyle{spr-mp-sola}
\bibliography{main.bib}  
%
%
%
%

\end{article} 
\end{document}